\def\CC{{\mathchoice
{\rm C\mkern-8mu\vrule height1.45ex depth-.05ex 
width.05em\mkern9mu\kern-.05em}
{\rm C\mkern-8mu\vrule height1.45ex depth-.05ex 
width.05em\mkern9mu\kern-.05em}
{\rm C\mkern-8mu\vrule height1ex depth-.07ex 
width.035em\mkern9mu\kern-.035em}
{\rm C\mkern-8mu\vrule height.65ex depth-.1ex 
width.025em\mkern8mu\kern-.025em}}}

\def\RR{{\rm I\kern-1.6pt {\rm R}}}

\def\ZZ{{\rm Z}\kern-3.8pt {\rm Z} \kern2pt}

\input phyzzx.tex
%\draft

\def\np{Nucl. Phys.}
\def\pl{Phys. Lett.}

\def\ap{Ann. Phys.}
\def\cmp{Comm. Math. Phys.}
\def\jmp{J. Math. Phys.}
\def\ijmp{Int. J. Mod. Phys.}
\def\mpl{Mod. Phys. Lett.}
\def\lmp{Lett. Math. Phys.}

\def\faa{Funct. Anal. and Appl.}
\def\pnas{Proc. Natl. Acad. Sci. USA}
\def\sjnp{Sov. J. Nucl. Phys.}
\def\nmt{Norsk. Mat. Tidsskrift}

\tolerance=500000
\overfullrule=0pt
\pubnum={US-FT-40/96\cr hep-th/9610224}
%\pubnum={US-FT-40/96}
\date={October, 1996}
\pubtype={}
\titlepage

\title{Structure constants for the ${\rm osp}(1\vert 2)$ 
 current algebra} 
\author{I.P. Ennes\foot{E-mail: ENNES@GAES.USC.ES}, 
A.V. Ramallo\foot{E-mail: ALFONSO@GAES.USC.ES} and 
J. M. Sanchez de Santos \foot{E-mail: SANTOS@GAES.USC.ES} }
\address{Departamento de F\'\i sica de
Part\'\i culas, \break Universidad de Santiago, \break
E-15706 Santiago de Compostela, Spain. \break}

\abstract{We study the free field realization of the 
two-dimensional ${\rm osp}(1\vert 2)$ current algebra. We
consider the case in which the level of the affine 
${\rm osp}(1\vert 2)$ symmetry is a positive integer. Using
the Coulomb gas technique we obtain integral representations
for the conformal blocks of the model. In particular, from
the behaviour of the four-point function, we extract the
structure constants for the product of two arbitrary primary
operators of the theory. From this result we derive the
fusion rules of the ${\rm osp}(1\vert 2)$ conformal field
theory and we explore the connections between the 
${\rm osp}(1\vert 2)$ affine symmetry and the $N=1$
superconformal field theories. }

\endpage
\pagenumber=1

\chapter{Introduction}

Current algebras play a prominent role in two-dimensional
Conformal Field Theory (CFT)
\REF\Review{For a review see J. Fuchs, ``Affine Lie algebras
and quantum groups", Cambridge University Press, 1992 and 
S. Ketov, ``Conformal Field Theory", World Scientific,
Singapore(1995).}[\Review]. In fact, the theories endowed
with an affine Kac-Moody symmetry can be considered as the
building blocks out of which most known CFT's can be
constructed. The basic procedures to generate new CFT's from
models possessing a current algebra symmetry are the coset
construction
\REF\GKO{P. Goddard, A. Kent and D.
Olive\journal\pl&B152(85)88\journal\cmp&103(86)105.}
[\GKO] and the hamiltonian reduction
\REF\BO{M. Bershadsky and H. Ooguri
\journal\cmp&126(89)49.}
[\BO]. Moreover, an
$sl(N)$ affine algebra shows up when quantum
two-dimensional $W_N$ gravity is analyzed in the light-cone
\REF\KPZ{A. M. Polyakov\journal\mpl&A2(87)893; V. G.
Knizhnik, A. M. Polyakov and A. B.
Zamolodchikov\journal\mpl&A3(88)819.}
gauge [\KPZ] and, as  was shown in refs. 
\REF\hu{H.L. Hu and M. Yu \journal\pl&B289(92)302
\journal\np&B391(93)389.}
\REF\yank{M. Spiegelglas and S. Yankielowicz
\journal\np&393(93)301;  O. Aharony et al.\journal
\np&B399(93)527\journal\pl&B289(92)309
\journal\pl&B305(93)35.}
[\hu, \yank], the topological
$sl(N)/sl(N)$ coset theories can be used to describe
non-critical $W_N$-strings.

One of the simplest Lie superalgebras is 
${\rm osp}(1\vert 2)$. In CFT the affine version of this
superalgebra is frequently encountered when one studies
models with $N=1$ supersymmetry. Indeed, the 
${\rm osp}(1\vert 2)$ current algebra is the starting point
in the construction of the $N=1$ superconformal minimal
models by means of the hamiltonian reduction procedure
\REF\bershadsky{M. Bershadsky and H.
Ooguri\journal\pl&B229(89)374.}
[\bershadsky]. It
also appears in the quantization of  two-dimensional
supergravity in the light-cone gauge
\REF\poly{A. M. Polyakov and A. B.
Zamolodchikov\journal\mpl&A3(88)1213.}
[\poly] and its topological
version, \ie\ the 
${\rm osp}(1\vert 2)/{\rm osp}(1\vert 2)$ coset model, is
related to the Ramond-Neveu-Schwarz non-critical
superstrings
\REF\yu{J. B. Fan and M. Yu, ``G/G Gauged
Supergroup Valued WZNW Field Theory", Academia Sinica preprint
AS-ITP-93-22, hep-th/9304123.}
\REF\Ennes{I. P. Ennes, J. M. Isidro and A. V.
Ramallo\journal\ijmp&A11(96)2379.}
[\yu, \Ennes]. It is thus interesting to have a clear
understanding of this  current algebra symmetry in order to
improve our knowledge of the two-dimensional superconformal
symmetry and, in particular, to unravel the connection
of the latter with the Lie superalgebra theory.

In this paper we shall determine the operator product algebra
of the primary fields of the ${\rm osp}(1\vert 2)$ current
algebra. With this purpose in mind, we shall develop a free
field representation of the ${\rm osp}(1\vert 2)$ theory,
which was previously introduced in ref. 
[\bershadsky] to study its
hamiltonian reduction.  These free field constructions have
become a powerful tool in CFT. Indeed, the Feigin-Fuchs
formalism
\REF\FF{B. L. Feigin and D. B. Fuchs
\journal\faa&13, No.4(79)91\journal\faa&16, No.2(82)47.}
[\FF], as  was spelt out by Dotsenko and Fateev in
ref. 
\REF\DF{Vl.S.Dotsenko and V. A. Fateev \journal\np&B240(84)312
\journal\np&B251(85)691\journal\pl&B154(85)291.}
[\DF], has allowed to represent the conformal blocks of
the minimal Virasoro models and to obtain their operator
algebra .  A similar analysis has been
performed in refs. 
\REF\SCFT{H. Eichenherr \journal\pl&B151(85)26; M.A. Bershadsky, 
V. G. Knizhnik and M. G. Teitelman \journal\pl&B151(85)31;
D. Friedan, Z. Qiu and S. Shenker \journal\pl&B151(85)37.}
\REF\Kita{Y. Kitazawa et al. \journal\np&B306(88)425.}
\REF\Zaugg{L. Alvarez-Gaume and P. Zaugg \journal\ap&215(92)171.}
[\SCFT, \Kita, \Zaugg] for the minimal superconformal models.

The free field representation of the bosonic current
algebras, the so-called Wakimoto representation, has been
introduced in refs.
\REF\waki{M. Wakimoto\journal\cmp&104(86)604.}
\REF\gera{A. Gerasimov et al. \journal\ijmp&A5(90)2495.}
\REF\Feigin{B. L. Feigin and E. V. Frenkel in ``Physics and
Mathematics of Strings", edited by L. Brink et al., World
Scientific, 1990 \journal\cmp&128(90)161 
\journal\lmp&19(90)307.} [\waki, \gera, \Feigin]. This
representation has been used by Dotsenko in ref.
\REF\Dotsenko{Vl. S.
Dotsenko\journal\np&B338(90)747\journal\np&B358(91)547.} 
 [\Dotsenko] to
evaluate the correlation functions and the structure constants
of the $sl(2)$ current algebra. In the ${\rm osp}(1\vert 2)$
case we shall follow the methodology of refs. 
[\DF, \Dotsenko]. We shall
restrict ourselves to the case in which the level of the
affine ${\rm osp}(1\vert 2)$ algebra is a positive integer.
This is equivalent in the
$sl(2)$ case to consider integrable representations. The
structure constants and fusion rules we have obtained have
been reported, without proof, in our previous paper
\REF\osp{I. P. Ennes, A. V. Ramallo and J. M. Sanchez de
Santos, ``On the free field realization of the 
${\rm osp}(1\vert 2)$ current algebra", Santiago preprint
US-FT-31/96, hep-th/9606180, to appear in Phys. Lett. B.}
[\osp]. We
present here a full account of our results.

The organization of the paper is as follows. In section 2 we
introduce the basic ingredients of our Coulomb gas
formalism. After recalling the free field representation of
the currents, which we take from ref. [\bershadsky], we
consider the realization of the primary fields. As  always
happens in this kind of free field realizations, we have two
representations for the same primary operator. In one of these
representations, which we shall refer to as the conjugate
representation, the unity is not realized as the trivial
operator. An analysis of the form of this conjugate unit
serves to establish the basic rules to compute vacuum
expectation values in the Fock space of our free field
representation. The last ingredient needed to compute
correlation functions is the screening charge. A local
operator, satisfying the requirements demanded to one of such
a charge, has been found in ref. [\bershadsky]. Once the basic
set-up of the formalism is in place, we can start to compute
the correlation functions of the theory. An important
consistency check of our approach is the verification that
our expectation values satisfy the selection rules dictated
by the ${\rm osp}(1\vert 2)$ representation theory, a brief
review of which is given in appendix A. At the end of section
2 we analyze the two- and three- point functions of the model
and we show that, indeed, our free field construction
incorporates the selection rules expected for the coupling of
${\rm osp}(1\vert 2)$ representations.

Section 3 is devoted to the study of the four-point
functions. By looking at the local behaviour of the conformal
blocks, we determine the possible s-channel intermediate
states and check that they correspond precisely to the
non-vanishing couplings found in the analysis of the
three-point functions. In the computation of the structure
constants we shall need the value of the normalization
integrals of the conformal blocks. The evaluation of these
integrals is  highly non-trivial and it is done in appendix B.

The determination of the operator algebra of the model is the
objective of section 4. We first construct the monodromy
invariant four-point correlators. Following the standard
procedure introduced in ref. [\DF] for the minimal models, the
structure constants can be obtained from the coefficients
appearing in the power expansion of the physical four-point
correlator. Actually, a suitable normalization must be
performed in order to properly identify  the structure
constants. After this is done, one is left with a long and 
uninspiring expression for these constants. It turns out,
however, that one can convert this expression into a symmetric
and rather transparent equation which is, actually, very
similar to the one found in refs.
\REF\ZF{A. B. Zamoloddchikov and V. A.
Fateev\journal\sjnp&43(86)637.}
 [\ZF, \Dotsenko] for the $sl(2)$ current
algebra.

The fusion rules that follow from our structure constants are
given in section 5. As we shall argue in this section, these
fusion rules provide some new insights on the relation
between the ${\rm osp}(1\vert 2)$ theory and the minimal
superconformal models, a fact  which was, actually,   one of
the main motivations for our work. We end this section with
some comments on how to extend our analysis to the case in
which $k$ is not integer. Finally, in section 6 we draw some
conclusions from our results and indicate some possible lines
of future research.

\chapter{Free field representation of the
 ${\rm osp}(1\vert 2)$  current algebra}

The  ${\rm osp}(1\vert 2)$ Lie superalgebra has 
three even (bosonic) generators and two odd (fermionic)
ones. Its affine version, \ie\ the  
${\rm osp}(1\vert 2)$  current algebra, is generated by
three bosonic and two fermionic currents. We shall denote
the former by $J_{\pm}$ and $H$ whereas for the latter we
shall use the symbols $j_{\pm}$. These currents can be
realized in terms of a scalar field $\phi$, a pair of
two conjugate bosonic fields $(w,\chi)$ and two
fermionic fields $\psi$ and $\bar\psi$. The
non-vanishing operator product expansions (OPE's) among
them will be taken as: 
$$
w(z_1)\,\chi(z_2)\,=\,\psi(z_1)\,\bar\psi(z_2)\,=\,{1\over z_1-z_2}
\,\,\,\,\,\,\,\,\,\,\,\,\,\,\,\,\,\,
\phi(z_1)\,\phi(z_2)\,=\,-{\rm log}\,(z_1-z_2)\,.
\eqn\uno
$$
The fields $\bar \psi$ and $\psi$ ($w$ and $\chi$)
constitute a $bc$ ($\beta\gamma$) system with conformal
dimensions $\Delta(\bar \psi)\,=\,1$ and 
$\Delta(\psi)\,=\,0$ ($\Delta(w)\,=\,1$ and
$\Delta(\chi)\,=\,0$ respectively). The explicit form of
the ${\rm osp}(1\vert 2)$ currents is [\bershadsky]:
$$
\eqalign{
J_+\,=&\,w\cr
J_-\,=&-\,w\chi^2\,+\,i\sqrt{2k+3}\,\,\chi\partial\phi\,-
\,\chi\psi\bar\psi\,+k\partial\chi\,+
\,(k+1)\psi\partial\psi\cr
H\,=&-w\chi\,+{i\over 2}\,\sqrt{2k+3}\,\,\partial\phi\,-\,
{1\over 2}\,\psi\bar\psi\cr
j_+\,=&\bar\psi\,+\,w\psi\cr
j_-\,=&-\chi(\bar\psi\,+\,w\psi)\,+i\sqrt{2k+3}\,\,
\psi\partial\phi\,+\,(2k+1)\partial\psi\,,\cr}
\eqn\dos
$$
where the c-number $k$ is the level of the affine 
${\rm osp}(1\vert 2)$ superalgebra. In what follows we
shall restrict ourselves to the case in which $k$ is a
positive integer. It may be easily verified using \uno\
that the currents $J_{\pm}$ and $H$ close an $sl(2)$
algebra. The fermionic operators $j_{\pm}$ extend this 
$sl(2)$ algebra to the full ${\rm osp}(1\vert 2)$
symmetry. In order to construct a CFT in which the
currents \dos\ are dimension-one primary fields, we must
first define  the energy-momentum tensor $T$ of the
model. For current algebras the Sugawara construction
provides a method to get $T$ as an expression quadratic
in the currents. In our case this construction yields: 
$$
T^J\,=\,{1\over
2k+3}\,[\,J_+J_-\,+\,J_-J_+\,+\, 2H^2\,-\,{1\over
2}\,j_+\,j_-\,+{1\over 2}\,j_-j_+\,]\,,
\eqn\tres 
$$
where normal-ordering is understood. Using standard
methods one can obtain $T$ as a function of our basic set
of free fields:
$$
T\,=\,w\partial\chi\,-\,\bar\psi\partial
\psi\,-\,{1\over 2}\,(\partial\phi)^2\,+\,
{i\over 2}\,\alpha_0\,\partial^2\phi\,.
\eqn\cuatro
$$
In eq. \cuatro\ $\alpha_0$ is a background charge for the
field $\phi$ which, in terms of the level $k$, can be
written as:
$$
\alpha_0\,=\,-{1\over \sqrt{2k+3}}\,.
\eqn\cinco
$$
It is straightforward to prove that the operator $T$
given in eqs. \tres\ and \cuatro\ satisfies the Virasoro
algebra with central charge given by:
$$
c\,=\,{2k\over 2k+3}\,.
\eqn\seis
$$

Let us construct now the primary fields of the theory. One
should have a multiplet of such fields associated to each 
irreducible representation of the superalgebra. The
representation theory of the ${\rm osp}(1\vert 2)$ Lie
superalgebra has been studied in ref.
\REF\Pais{A. Pais and V. Rittenberg\journal\jmp&16(75)2063; 
M. Scheunert, W. Nahn and  V. Rittenberg\journal\jmp&18(77)155.}
\REF\scheunert{M. Scheunert,  ``The Theory of Lie
Superalgebras",  {\sl Lect. Notes in Math.} 716,
Springer-Verlag, Berlin (1979) and  
L. Frappat, P. Sorba and A. Sciarrino, ``Dictionary on
Lie Superalgebras", hep-th/9607161.}
 [\Pais](see also ref. 
[\scheunert] for an account of the general theory of Lie
superalgebras).
Let us recall some of its basic features. A more detailed
review is given in  appendix A. The ${\rm osp}(1\vert 2)$
finite dimensional irreducible representations closely
resemble those of the $sl(2)$ algebra. They are
characterized by an integer or half-integer number $j$ which
we shall refer to as the isospin of the representation. A
general state of the isospin $j$ representation will be denoted
by $|j,m>$, where $m$ is the eigenvalue of the Cartan
generator $H$ 
($m=-j, -j+{1\over 2},\,\cdots\,,j-{1\over 2}, j$). Acting with
the odd generators $j_{\pm}$, the value of $m$ is shifted by
$\pm 1/2$, while the bosonic currents $J_{\pm}$ produce a
change in the eigenvalue  $m$ of one unit. The highest weight
state of the isospin $j$ representation, whose dimensionality
is  $4j+1$, is $|j,j>$. To completely characterize 
the representation we must specify, in addition, the
statistics of its highest weight state. If $|j,j>$ is bosonic
(fermionic) we will say that the representation is even (odd).
It is important to point out that when $j-m$ is integer
(half-integer) the states $|j,m>$ and $|j,j>$ have the same
(opposite) statistics. 

The primary field associated to the $|j,m>$ state will be
denoted by $\Phi^j_m$. The conformal dimensions $\Delta_j$ of
these operators can be written in terms of the quadratic
Casimir invariant $C_2$ of the ${\rm osp}(1\vert 2)$
superalgebra. The expression of $C_2$ is given in  appendix
A. By comparing this expression with the bilinear form
appearing in $T$ (see eq. \tres) one easily concludes that:
$$
\Delta_j\,=\,{2C_2\over 2k+3}\,.
\eqn\siete
$$
Taking into account that for an isospin $j$ representation the
value of $C_2$ is $j(j+{1\over 2})$ (see eq. (A5)), one can
rewrite eq. \siete\ as:
$$
\Delta_j\,=\,{j(2j+1)\over 2k+3}\,.
\eqn\ocho
$$

The degenerate admissible representations of 
${\rm osp}(1\vert 2)$ have been studied in refs.
\REF\KW{V. Kac and M. Wakimoto\journal\pnas&85(88)4956.}
\REF\yudos{J. B. Fan and M. Yu, ``Modules over affine Lie
superalgebras", Academia Sinica preprint AS-ITP-93-22.}
 [\KW, \yudos] by
considering the coset decomposition of 
${\rm osp}(1\vert 2)$ into the $sl(2)$ algebra generated
by its bosonic currents and an 
${\rm osp}(1\vert 2)/sl(2)$ theory. It was shown in refs.
[\KW, \yudos] that when the ${\rm osp}(1\vert 2)$  level $k$
is  admissible with respect to the even $sl(2)$ algebra, 
 the ${\rm osp}(1\vert 2)/sl(2)$ theory can be
identified with one of the models of the Virasoro
minimal series. The ${\rm osp}(1\vert 2)$ levels $k$ for
which degenerate representations appear are parametrized
by two integers $p$ and $q$, by means of the relation 
$2k+3={q\over p}$, where $q+p$ is even and $p$ and 
${p+q\over 2}$ are relatively prime. The isospins $j$
corresponding to the admissible representations are
determined by the equation 
$4j+1=r-s{q\over p}$ with $r+s$ odd and $r,s$ taking
values in the ranges $r=1,\cdots,q-1$ and
$s=0,\cdots,p-1$. In our case, \ie\ when $k$ is a positive
integer, $p=1$ and thus $s=0$. Therefore, since
$r$ must be odd, the
highest value it can take is $2k+1$ and,  thus, we conclude 
that the admissible representations have integer and
half-integer isospins
$j$ that satisfy $j\le k/2$. Notice that this corresponds
to taking integral representations for the even $sl(2)$
algebra. It will be understood from now on that this
constraint is satisfied by all the primary fields 
$\Phi_m^j$ we shall consider.

The actual form of the operators $\Phi^j_m$ in our free field
realization can be determined
as follows. First of all, it is clear that one can obtain the
fields $\Phi^j_m$ with $m<j$ by acting with the lowering
operators $J_-$ and $j_-$ on the highest weight field 
$\Phi^j_j$. Moreover,  the OPE of the raising currents 
$J_+$ and $j_+$ with $\Phi^j_j$ 
must vanish. By inspecting the realization of 
$J_+$ and $j_+$ in eq. \dos, one immediately reaches the
conclusion that in the expression of $\Phi^j_j$ only the
fields $w$ and $\phi$ can appear. Let us suppose that we
adopt an ansatz  for $\Phi^j_j$ in which the field $w$ is not
present. We therefore shall assume that $\Phi^j_j$ can be
written as a vertex operator of the form: 
$$
\Phi^j_j\,=\,e^{i\alpha_j\,\alpha_0\,\phi},
\eqn\nueve
$$
where $\alpha_j$ is a constant whose exact dependence on $j$
has to be determined. The easiest way to fix the
value of $\alpha_j$ is by requiring that the current $H$ acts 
diagonally on $\Phi^j_j$ with eigenvalue equal to $j$. The
$H$-charge of the operator \nueve\ is $-\alpha_j/2$, as an
straightforward calculation using eqs. \dos\ and \cinco\ shows
and, thus, we must take $\alpha_j=-2j$. It is also simple to
verify that for this value of $\alpha_j$ the conformal
dimension of $\Phi^j_j$ is given precisely by eq. \ocho. As
we have pointed out above, the other members $\Phi^j_m$ of the
field multiplet can be obtained by successive application of
the operators $J_-$ and $j_-$. The result one arrives at is
the following:
$$
\Phi^j_m\,=\,\cases{\chi^{j-m}\,e^{-2ij\alpha_0\,\phi}
                     &if $j-m\in \ZZ$\cr\cr
                    \chi^{j-m-{1\over 2}}\,\psi\,
                    e^{-2ij\alpha_0\,\phi}
                    &if $j-m\in \ZZ\,+{1\over 2}\,$.}
\eqn\diez
$$
Notice that the conformal dimensions of the operators \diez\
are $m$-independent and given by eq. \ocho( the fields $\chi$
and $\psi$ have vanishing conformal weight). It is also easy
to obtain the action of the currents on the operators \nueve.
The OPE of the Cartan current $H$ and the fields \diez\ is:
$$
H(z_1)\,\Phi^j_m(z_2)\,=\,m\,\,
{\Phi^j_{m}(z_2)\over z_1-z_2}\,,
\eqn\once
$$
which confirms that the integer $m$ in eq. \diez\ is the
$H$-charge. Moreover,  the operators $J_{\pm}$ connect two
fields whose value of $m$ differs in one unit:
$$
J_{\pm}(z_1)\,\Phi^j_m(z_2)\,=\,\cases{
             (j\mp m)\,\,{\Phi^j_{m\pm 1}(z_2)\over z_1-z_2}
             &if $j-m\in \ZZ$\cr\cr\cr
             (j\mp m\,-{1\over 2})\,\,{\Phi^j_{m\pm 1}(z_2)
              \over z_1-z_2}
             &if $j-m\in \ZZ\,+{1\over 2}\,$.}
\eqn\doce
$$
Finally, as they should, the fermionic currents $j_{\pm}$
change the value of $m$ by one-half unit:
$$
j_{\pm}(z_1)\,\Phi^j_m(z_2)\,=\,\cases{
             (j\mp m)\,\,{\Phi^j_{m\pm 1/2}(z_2)\over z_1-z_2}
             &if $j-m\in \ZZ$\cr\cr\cr
             \pm\,\,{\Phi^j_{m\pm 1/2}(z_2)\over z_1-z_2}
             &if $j-m\in \ZZ\,+{1\over 2}\,$.}
\eqn\trece
$$

The representation \diez\ of the primary fields is not unique.
Indeed, as we will show below, one can find a representation
for these fields which is conjugate to the one written in \diez.
Recall that in  \nueve\ we have discarded the
possibility of having a power of the field $w$. Let us now
include this type of term in our ansatz for the highest weight
operator. For reasons that soon will become apparent, it is
convenient to find first the conjugate of the isospin zero
operator. Let us call $I$  the operator \diez\ for $j=m=0$.
It is obvious from \diez\ that $I=1$, so we are trying to get
a new representation for the unit operator. This conjugate unit
will be denoted by $\tilde I$. Its expression will be of the
form:
$$
\tilde I\,=\,w^A\,e^{iB\alpha_0\,\phi}\,,
\eqn\catorce
$$
where $A$ and $B$ are constants. As we have already discussed,
the highest weight conditions:
$$
j_{+}\,(z_1)\,\tilde I(z_2)\,=\,
J_{+}\,(z_1)\,\tilde I(z_2)\,=\,0\,,
\eqn\quince
$$
are satisfied for any value of $A$ and $B$. Moreover, the
singular terms in the product of the current $H$ and 
$\tilde I$ are given by:
$$
H(z_1)\,\tilde I(z_2)\,=\,(\,A-{1\over 2}\,B\,)\,
{\tilde I\,(z_2)\over z_1-z_2}\,.
\eqn\dseis
$$
Therefore, if we require neutrality of $\tilde I$ with respect
to $H$ (recall that $\tilde I$ corresponds to $m=0$), the
following condition must be imposed
$$
B\,=\,2A\,.
\eqn\dsiete
$$
Eq. \dsiete\ fixes $B$ in terms of $A$. This latter constant
can be determined by looking at the current algebra
descendants of the conjugate unit $\tilde I$. In fact, acting
with 
$j_{-}$ and $J_{-}$ on  $\tilde I$, one gets:
$$
\eqalign{
j_{-}\,(z_1)\,\tilde I(z_2)\,=&\,
{\xi\,(z_2)\over z_1-z_2}\cr
J_{-}\,(z_1)\,\tilde I(z_2)\,=&\,
A(k+A+1)\,
{[w(z_2)]^{A-1}\,\,e^{2iA\alpha_0\phi(z_2)}
\over (z_1-z_2)^2}\,\,+\,\,
{\Xi\,(z_2)\over z_1-z_2}\,,\cr}
\eqn\docho
$$
where $\xi\,(z)$ and $\Xi\,(z)$ are operators whose precise
form will not be needed. One should not have double pole
singularities in the OPE's of $j_{-}$ and $J_{-}$ with a
highest weight operator. These double pole singularities do
not appear in \docho\ when $A=-k-1$. Moreover, one can easily
verify that, apart from the trivial solution $A=0$, only for
this value of
$A$ the conformal dimension of $\tilde I$ is zero. We are thus
led to adopt the following expression for $\tilde I$:
$$
\tilde I\,=\,w^s\,e^{2is\alpha_0\phi}\,,
\eqn\dnueve
$$
where
$$
s\,=\,-k-1\,.
\eqn\veinte
$$

Similarly to what happens in the $sl(2)$ case [\Dotsenko], one
can show that the descendant fields $\xi$ and $\Xi$ generate
null vectors in the module of the trivial $j=0$ representation.
Indeed, one can check that $\xi$ and $\Xi$ satisfy the highest
weight conditions for the ${\rm osp}(1\vert 2)$ current
algebra and, thus, they are going to decouple (\ie\ to vanish)
in the conformal blocks of the model. Moreover, the form
\dnueve\ of the conjugate identity fixes the charge asymmetry of
the Fock space metric. This charge asymmetry is present in
other Coulomb gas representations of CFT's
[\DF, \SCFT]. The best way to
determine it is by requiring  the vacuum expectation value
of the conjugate identity to be non-vanishing. By inspecting
\dnueve, one can easily obtain a series of   selection rules
that the non-vanishing correlators must satisfy . Let us imagine
that we are computing the expectation value 
$<\,\prod_{i}\,O_i\,>$, where $O_i$ are general operators of
the form 
$O_i\,=\,w^{n_i}\,\chi^{m_i}\,e^{i\alpha_i\phi}\,$. Calling 
$N(w)\,=\,\sum_i\,n_i$ and $N(\chi)\,=\,\sum_i\,m_i$, one gets
the following conditions:
$$
\eqalign{
&N(w)\,-\,N(\chi)\,=\,s\cr
&\sum_i\,\alpha_i\,=\,2\alpha_0 s\,.\cr}
\eqn\vuno
$$
A possible way to implement the conditions \vuno\ is by
defining the  ${\rm osp}(1\vert 2)$  correlators with some
fields inserted at the point at infinity. We will not need,
however,  to be very explicit about this point as far as the
conditions \vuno\ are satisfied. Notice that when $k$ is a
positive integer, $s$ is negative. Therefore, in some of the
expressions of the conjugate fields (such as the one of
$\tilde I$ in \dnueve) we are going to have negative powers of
the $w$ field. It will be understood in what follows that
these negative powers have been properly defined (the
situation is similar to the $sl(2)$ case, see [\Dotsenko]).

For $j>0$ the form of the conjugate operators can be obtained
in a similar way. In general, the highest weight operator 
$\tilde \Phi_j^j$ will be given by an expression of the type 
 \catorce. Requiring now the $H$-eigenvalue to be $j$, one
gets the constraint $A-{B\over 2}\,=\,j$, whereas by fixing
the conformal dimension of $\tilde \Phi_j^j$ to the value
\ocho\ one obtains $A=2j+s$ and $B=2j+2s$. Therefore 
$\tilde \Phi_j^j$ is given by:
$$
\tilde \Phi_j^j\,=\,w^{2j+s}\,\,
e^{2i(j+s)\alpha_0\,\phi}\,.
\eqn\vdos
$$
The remaining operators $\tilde \Phi_m^j$ of the conjugate
multiplet can be obtained by acting with the currents $j_-$
and $J_-$ on the highest weight $\tilde \Phi_j^j$. After a
simple calculation one gets:
$$
\eqalign{
j_{-}(z_1)\,\tilde \Phi_{j}^j\,(z_2)\,=&\,2j\,
{\tilde\Phi_{j-1/2}^{j}(z_2)\over z_1-z_2}\cr
J_{-}(z_1)\,\tilde \Phi_j^j\,(z_2)\,=&\,
{\tilde\Phi_{j-1}^{j}(z_2)\over z_1-z_2}\,,\cr}
\eqn\vtres
$$
where $\tilde\Phi_{j-1/2}^{j}$ is given by:
$$
\tilde \Phi_{j-1/2}^j\,=\,{1\over 2j}\,[\,
(2j+s)\,\bar\psi\,w^{2j+s-1}\,-\,s\,w^{2j+s}\,\psi\,]\,
e^{2i(j+s)\,\alpha_0\phi}\,,
\eqn\vcuatro
$$
and $\tilde\Phi_{j-1}^{j}$ can be written as:
$$
\eqalign{
\tilde\Phi_{j-1}^{j}\,=&\,\Bigl[\,\,\chi\,\omega^{2j+s}\,-\,
{(2j+s)(2j+s-1)\over 2j}\,\,\partial\omega\,
\omega^{2j+s-2}\,+\cr
&+\,{2j+s\over 2j}\,[\,\psi\bar\psi\,-\,i\,
\sqrt{2k+3}\,\,\partial\phi\,]\,\omega^{2j+s-1}\,\,
\Bigr]\,e^{2i(j+s)\,\alpha_0\phi}\,.
\cr}
\eqn\vcinco
$$
Notice that the OPE's  \vtres\  are the same as those of
$j_-$ and $J_-$ with the fields \diez (see eqs. \doce\ and
\trece). This fact can be confirmed by computing other
singular product expansions such as:
$$
\eqalign{
j_{\pm}(z_1)\,\tilde \Phi_{j-1/2}^j\,(z_2)\,=&\,\pm 2j\,
{\tilde\Phi_{j-1/2\pm 1/2}^{j}(z_2)\over z_1-z_2}\cr
j_{+}(z_1)\,\tilde \Phi_{j-1}^{j}\,(z_2)\,=&\,
{\tilde\Phi_{j-1/2}^{j}(z_2)\over z_1-z_2}\,\,.\cr}
\eqn\vseis
$$
By successive application of the currents $j_-$ and $J_-$, one
can generate other components of the conjugate multiplet of
primary fields. In general, the expressions of $\tilde
\Phi_m^j$  found in this way are increasingly complicated as
$m$ is decreased. Fortunately for our purposes  the
knowledge of the highest weight conjugate field $\tilde
\Phi_j^j$ will be enough.

Following the standard procedure to compute correlators in the
Coulomb gas realizations of CFT's, one must represent the
conformal blocks of the theory as expectation values of
products of the fields $\Phi_{m}^{j}$ and their conjugates 
$\tilde\Phi_{m}^{j}$. However, in order to get a non-vanishing
result, the conditions \vuno\ must be fulfilled. This can only
be achieved if the fields are screened by means of the
insertion of a new operator $Q$ (the screening charge), which
must be invariant under the action of the 
${\rm osp}(1\vert 2)$ currents and must have zero conformal
dimension. We shall represent $Q$ as an integral of a local
operator $S(z)$ over a closed contour:
$$
Q\,=\,\oint\,dz\,S(z)\,,
\eqn\vsiete
$$
where $S(z)$ has conformal weight equal to one and is such
that its OPE's with the ${\rm osp}(1\vert 2)$ currents have
only total derivatives. This last condition guarantees the
(anti)commutation of $Q$ with $J_{\pm}$, $H$ and $j_{\pm}$. As
it has been checked in ref. [\bershadsky], $S$ can be taken as:
$$
S\,=\,(\,\bar\psi\,-\,w\psi\,)\,e^{i\alpha_0\phi}\,.
\eqn\vocho
$$

In the remaining of this section we shall study the two- and
three-point functions of the model. We shall verify that the
formalism we have introduced correctly reproduces  the features
of these correlators which are to be expected from the
conformal invariance and the ${\rm osp}(1\vert 2)$
representation theory (see appendix A). 

The two-point function
can be represented as an expectation value of the product of a
field \diez\ and its conjugate. It can be easily seen that the
conditions \vuno\ can be satisfied without the insertion 
of the screening charge $Q$. Let us check this fact in the
simplest case in which the highest weight conjugate operator
\vdos\ is used. The expectation value to be computed is:
$$
<\,\Phi^{j}_{-j}(z_1)\,\tilde\Phi^{j}_{j}(z_2)\,>\,=\,
<\,[\chi(z_1)]^{2j}\,e^{-2ij\alpha_0\phi(z_1)}\,
[w(z_2)]^{2j+s}\,e^{2i(j+s)\alpha_0\phi(z_2)}\,>\,.
\eqn\vnueve
$$
A simple counting shows that, indeed, the conditions \vuno\
are satisfied. Moreover, an elementary application of Wick's
theorem allows to write the right-hand side of eq. \vnueve\
as:
$$
<\,\Phi^{j}_{-j}(z_1)\,\tilde\Phi^{j}_{j}(z_2)\,>\,=\,
{C\over (z_1-z_2)^{2\Delta_j}}\,,
\eqn\treinta
$$
where $\Delta_j$ is given in \ocho\ and $C$ is a constant
proportional to the expectation value of  $w^s$ in the 
${\rm osp}(1\vert 2)$ Fock space. One can also verify that for
other  $H$-eigenvalues of the fields, such as in 
$<\,\Phi^{j}_{-j+{1\over 2}}(z_1)\,
\tilde\Phi^{j}_{j-{1\over 2}}(z_2)\,>$, the correlator is also
given by the right-hand side of \treinta\ (with the same value
of the constant $C$). This is precisely the behaviour expected
for an ${\rm osp}(1\vert 2)$ CFT.

The three-point conformal blocks can be represented as
correlators of two fields \diez\ and one conjugate operator
$\tilde\Phi^{j}_{m}$. In
this case screening charges must be inserted in order to
satisfy the ${\rm osp}(1\vert 2)$ charge asymmetry conditions.
Therefore we must consider an expectation value of the form:
$$
<\,\Phi^{j_1}_{m_1}(z_1)\,\Phi^{j_2}_{m_2}(z_2)\,
\tilde\Phi^{j_3}_{m_3}(z_3)\,Q^n\,>\,.
\eqn\tuno
$$
It is a simple exercise to determine the number $n$ of
screening charges needed in order to  have a non-zero
result. Although we have not obtained the explicit expression
of $\tilde\Phi^{j}_{m}$ for arbitrary $m$, 
we do know  that the
coefficient of the  $\phi$ field in the exponential is not
changed by the action of the $j_-$ and $J_-$ currents and, 
therefore,  for all values of $m$ this coefficient is the same
as in eq. \vdos. Thus we can write the value of the left-hand
side of the second equation in \vuno\ for the correlator \tuno\
as:
$$
\sum\alpha_i\,=\,2\alpha_0\,(\,s+j_3-j_1-j_2+{n\over 2}\,)\,.
\eqn\tdos
$$
Therefore from \vuno\ we obtain that $n$ is related to $j_1$,
$j_2$ and $j_3$ through the expression:
$$
0\le j_1+j_2-j_3\,=\,{n\over 2}\,,
\eqn\ttres
$$
where the inequality is an obvious consequence of the fact
that $n$ is a positive integer. Eq. \ttres\ implies, in
particular, that $j_3\le j_1+j_2$. Moreover, the three-point
function we are considering could equally be represented by
taking the conjugate operator to be the one of isospin $j_1$
or that of isospin $j_2$. In these two cases from the screening
condition we get two inequalities similar to \ttres, namely:
$$
\eqalign{
&j_1+j_3-j_2\,\ge\,0\cr
&j_2+j_3-j_1\,\ge\,0\,.\cr}
\eqn\tcuatro
$$
Notice that, combining the two inequalities in \tcuatro, we get
that $j_3\,\ge\,|\,j_1-j_2\,|$. Taking  \ttres\ into account,
we get that the values of $j_3$ that can have a non-vanishing
coupling to the isospins $j_1$ and $j_2$ are contained in the
interval:
$$
|\,j_1-j_2\,|\le\,j_3\,\le\,j_1+j_2\,.
\eqn\tcinco
$$
The similarity of eq. \tcinco\ with what happens in $sl(2)$ is
manifest. It is important to point out, however, that in the 
${\rm osp}(1\vert 2)$ case $j_1+j_2-j_3$ is in general
half-integer (see eq. \ttres). These  are well-known 
results of the ${\rm osp}(1\vert 2)$  representation theory. The
fact that we were able to reproduce them within our Coulomb
gas formalism is a confirmation of the correctness of our
approach. As a further verification, let us check that the
expectation value \tuno\ is non-vanishing only when 
$m_1+m_2+m_3=0$. To prove this statement the first of our
selection rules \vuno\ will become crucial. Actually, we are
only going to consider the simplest case in which $m_3=j_3$.
Thus, we will be dealing  with the correlator:
$$
<\,\Phi^{j_1}_{m_1}(z_1)\,\Phi^{j_2}_{m_2}(z_2)\,
\tilde\Phi^{j_3}_{j_3}(z_3)\,Q^n\,>\,.
\eqn\tseis
$$
The screening charge $Q$ is the sum of two terms, and only
in one of them  the $w$ field is present. Therefore in $Q^n$
there will be contributions with different powers $w^l$ with 
$0\le l\le n$. Only one of these terms, \ie\ the one that
satisfies \vuno, contributes to the correlator \tseis.
Therefore, inside this expectation value we can
 substitute:
$$
Q^n\,\sim\,\oint\,w^l\,\psi^l\,\bar\psi^{\,n-l}\,,
\eqn\tsiete
$$
where a symbolic notation for the multiple contour integral
has been adopted. Let us now determine the value of
$l$ in \tsiete. Our first observation is that the number of
$\chi$ fields in the operator $\Phi^{j}_{m}$ is equal to the
integer part of $j-m$, which we shall denote by  $[j-m]$ (see
eq. \diez).  Therefore for the correlator \tseis\ one has:
$$
N(w)\,-\,N(\chi)\,=\,2j_3+s+l-[j_1-m_1]-[j_2-m_2]\,.
\eqn\tocho
$$
After taking into account \vuno, we conclude that $l$ is given
by:
$$
l\,=\,[j_1-m_1]+[j_2-m_2]-2j_3\,.
\eqn\tnueve
$$
The number of $\psi$'s and $\bar\psi$'s inside a correlator
must be equal if we want to have a chance of getting a
non-vanishing result. It is straightforward to prove that the
number of  $\psi$ fields in the operator $\Phi^{j}_{m}$ is 
$2(\,j-m-[j-m]\,)$. Since $\tilde\Phi^{j}_{j}$ does not
contain any $\psi$ field in its expression (see eq. \vdos),
the number of $\psi$'s in the non-vanishing terms of \tseis\
 is:
$$
l+2(\,j_1-m_1+j_2-m_2-[j_1-m_1]-[j_2-m_2]\,)\,.
\eqn\cuarenta
$$
The only source of  $\bar\psi$'s in \tseis\ is the screening
charge $Q$. After inspecting eq.  \tsiete, we conclude that
their number in the correlator is $n-l$. Taking into account
the values of $n$ and $l$ given in \ttres\ and \tnueve, it is
easy to prove that \cuarenta\ is equal to $n-l$ only when 
$m_1+m_2+j_3\,=\,0$. This is the result we wanted to
demonstrate.

\chapter{The four-point functions}

In this section we shall apply the free field representation
studied in section 3 to the computation of the four-point
conformal blocks of the model. As in the case of the two- and
 three-point functions, we shall represent these blocks as
expectation values of primary fields in the 
${\rm osp}(1\vert 2)$ Fock space. In order to satisfy the
charge asymmetry conditions of the latter, one of the
four primary fields will be taken in the conjugate
representation. Therefore we shall consider a correlator of the
form 
$<\Phi^{j_1}_{m_1}(z_1)\,\Phi^{j_2}_{m_2}(z_2)\,
\Phi^{j_3}_{m_3}(z_3)\,\tilde\Phi^{j_4}_{m_4}(z_4)\,
Q^n\,>$. The number $n$ of screening charges can be easily
determined from the second condition in \vuno. Indeed, one can
immediately demonstrate that only when 
$n\,=\,2\,(\,j_1\,+\,j_2\,+\,j_3\,-\,j_4\,)$ this correlator
is non-vanishing. By using the $sl(2)$ projective
invariance of the Virasoro algebra, we can fix the positions
of the four fields to the values $z_1=0$, $z_2=z$, $z_3=1$ and
$z_4=\infty$. After this fixing, the correlator is a function
of the variable $z$. We want to investigate the analytical
structure of these blocks and, as a result of this study, we
would like to determine the operator algebra of the model. We
shall follow the method developed in ref. [\DF] for the minimal
models and extended to $sl(2)$ 
current algebras in ref. [\Dotsenko]. As
it happened in this latter case, the study of the
correlator for some particular values of the isospins and
$H$-charges is enough to determine the structure constants of
the model. Therefore, as 
in ref. [\Dotsenko], we shall restrict
ourselves to the situation in which $j_3=j_2$ and $j_4=j_1$ with
$j_1\ge j_2$. Notice that in this case $n\,=\,4j_2$. This
implies that the number of screening operators must be even.
In addition, the $H$-charges of the four primary fields will be
taken to be $m_1\,=-j_1$, $m_2\,=\,j_2$, $m_3\,=-j_2$ 
and $m_4\,=\,j_1$. Therefore we will center our efforts in the
analysis of the quantity:
$$
I(z)\,\equiv\,
<\,\Phi^{j_1}_{-j_1}(0)\,\Phi^{j_2}_{j_2}(z)\,
\Phi^{j_2}_{-j_2}(1)\,\tilde\Phi^{j_1}_{j_1}(\infty)\,
Q^{4j_2}\,>\,.
\eqn\cuno
$$
We shall suppose, finally, that the four representations
involved in the correlator \cuno\ are even, \ie\ with bosonic
highest weights. As  is reviewed in  appendix A, for
these even representations one can choose a metric such that
all the states of the corresponding multiplet have positive
norm. 

Using the expressions of the primary fields and the screening
charge, it is immediate to get the explicit representation of
$I(z)$. One has:
$$
I(z)\,=\,
\prod_{i=1}^{n}\,\,\oint_{C_i}\,\,d\tau_i\,
\lambda(z,\{\tau_i\})\,\eta(\{\tau_i\})\,.
\eqn\cdos
$$
In eq. \cdos\ $\tau_i$ are the integration variables that
appear in the screening charges, the integration contours
$C_i$ will be specified below and the function
$\lambda(z,\{\tau_i\})$ is the part of the correlator that
corresponds to the field $\phi$, namely:
$$
\eqalign{
\lambda(z,\{\tau_i\})\,=\,
<\,e^{-2ij_1\alpha_0\,\phi(0)}&\,e^{-2ij_2\alpha_0\,\phi(z)}\,
e^{-2ij_2\alpha_0\,\phi(1)}\,
e^{2i(s+j_1)\alpha_0\,\phi(\infty)}\,\times\cr
&\times e^{i\alpha_0\,\phi(\tau_1)}\cdots
 e^{i\alpha_0\,\phi(\tau_n)}\,>\,.\cr}
\eqn\ctres
$$
The function $\eta(\{\tau_i\})$ contains the 
contribution of
the fields $w$, $\chi$, $\psi$ and $\bar \psi$ to the vacuum
expectation value \cuno.  
Notice that one gets many terms in 
$\eta(\{\tau_i\})$ when $Q^{4j_2}$ is expanded as in eq.
\tsiete (with $n=4j_2$). Actually, only one type of these 
terms is non-vanishing. Indeed, as in the primary
fields involved in \cuno, the fields $\psi$ and $\bar\psi$ do
not appear, only those pieces of $Q^{4j_2}$ with equal number
of $\psi$'s and $\bar\psi$'s survive in the expectation value
\cuno. Inspecting eq. \tsiete\ we conclude that those
contributions appear  when $l=n/2=2j_2$. Therefore we can
write:
$$
\eqalign{
\eta(\{\tau_i\})\,=&\,(-1)^{2j_2}\,
<\,(\chi(0))^{2j_1}\,(\chi(1))^{2j_2}\,
(w(\infty))^{2j_1+s}\,w(\tau_1)\,\cdots\,w(\tau_{2j_2})\,\,>\times\cr\cr
&\times\,<\,\psi(\tau_1)\cdots\psi(\tau_{2j_2})\,
\bar\psi(\tau_{2j_2+1})\cdots\bar\psi(\tau_{4j_2})\,>\,+\,
{\rm permutations.}\cr}
\eqn\ccuatro
$$
In eq. \ccuatro\ the sum over permutations has its
origin in all the 
possible terms of the type \tsiete\ with $l=2j_2$. Notice
that in eq. \ccuatro\ $N(w)=2j_1+2j_2+s$ and
$N(\chi)=2j_1+2j_2$ and thus the selection rule \vuno\ is
satisfied.

We will use in eq. \cdos\ the canonical set of contours that
give rise to the s-channel 
conformal blocks [\DF, \Dotsenko]. Notice that
our correlators \cuno\ have $z_1=0$, $z_2=z$, $z_3=1$ and
$z_4=\infty$ as singular points. The $C_i$'s will be contours
connecting these points as follows. We shall take the first
$n-p+1$ integrals along a path lying on the real axis and
joining the points $\tau=1$ and $\tau=\infty$. The remaining
$p-1$ integrals, \ie\ those involving the variables 
$\tau_{n-p+1+i}$ for $i=1,\cdots, p-1$, will be taken
along the segment $(0,z)$. Obviously the range of values of
$p$ is $1\le p\le 4j_2+1$. We have thus divided our
integrations in two sets : $n-p+1$ of them are performed in
the interval $(1,\infty)$ while for the other $p-1$ 
the domain of integration is the segment
that goes from $\tau=0$ to $\tau=z$. Within each of these two
intervals the integration variables will be taken as ordered.
Thus, if we relabel the $\tau_i$'s as 
$u_i=\tau_i$ for $i=1,\cdots, n-p+1$ and $v_i=\tau_{n-p+1+i}$
for  $i=1,\cdots, p-1$, the conformal block $I_p(z)$ can be
written as:
$$
I_p(z)\,=\,\int_1^{\infty}\,du_1\cdots\int_1^{u_{n-p}}\,
du_{n-p+1}\int_0^{z}\,dv_1\cdots\int_0^{v_{p-2}}\,dv_{p-1}\,
\lambda_p(z,\{u_i\},\{v_i\})\,\eta_p(\{u_i\},\{v_i\}).
\eqn\ccinco
$$
In eq. \ccinco\ the quantities $\lambda_p(z,\{u_i\},\{v_i\})$
and $\eta_p(\{u_i\},\{v_i\})$ are, respectively,  the functions 
$\lambda(z,\{\tau_i\})$ and  $\eta(\{\tau_i\})$ after the
relabelling of variables introduced above. By applying Wick's
theorem to the vacuum expectation value \ctres, one can
readily  prove that  $\lambda_p(z,\{u_i\},\{v_i\})$ 
is given by:
$$
\eqalign{
\lambda_p(z,\{u_i\},\{v_i\})\,=&\,z^{8j_1j_2\rho}\,
(1-z)^{8j_2^2\rho}\,
\prod_{i=1}^{n-p+1}\,u_i^a\,(u_i-z)^b\,(u_i-1)^b\,
\prod_{i<j}(u_i-u_j)^{2\rho}
\times\cr
&\times
\prod_{i=1}^{p-1}\,v_i^a\,(z-v_i)^b\,(1-v_i)^b
\prod_{i<j}(v_i-v_j)^{2\rho}
\,\prod_{i=1}^{n-p+1}\,\prod_{j=1}^{p-1}
(u_i-v_j)^{2\rho},\cr}
\eqn\cseis
$$
where
$$
\rho\,=\,\alpha_0^2/2\,=\,{1\over 2(2k+3)}\,,
\eqn\csiete
$$
and the constants $a$ and $b$ are defined as:
$$
a\,=\,-2j_1\alpha_0^2\,,
\,\,\,\,\,\,\,\,\,\,\,\,\,\,
b\,=\,-2j_2\alpha_0^2\,.
\eqn\cocho
$$

The representation \ccinco\ can be used to obtain the
non-analytic behaviour  of the blocks around the point $z=0$.
In general, one expects that, as $z\rightarrow 0$
$$
I_p(z)\,\sim\,N_p\,z^{\gamma_p}\,,
\eqn\cnueve
$$
where $N_p$ and $\gamma_p$ are constants and only the leading
term of the expansion has been written down. Eq. \cnueve\
corresponds to a well-defined s-channel exchange. The
exponents $\gamma_p$ are related to the conformal weights of
the s-channel intermediate states, whereas the $N_p$'s measure
the coupling constants of the intermediate channel and are
related to the structure constants of the operator algebra of
the model (see below). In order to make the $z\sim 0$ behaviour
more explicit let us rescale the $v_i$ integration variables
as $v_i\,=z\,t_i$.  These  new variables $t_i$ are
integrated over the interval $(0,1)$. It is an easy exercise
to obtain the leading term of the $z\rightarrow 0$ expansion
of $\lambda_p(z,\{u_i\},\{zt_i\})$. One has:
$$
\eqalign{
\lambda_p(z,\{u_i\},\{zt_i\})\,\sim\,\,&
\,z^{8j_1j_2\rho+(p-1)\,[\,a+b+(p-2)\rho\,]}\,
\prod_{i=1}^{n-p+1}\,u_i^{a+b+2\rho \,(p-1)}
\,(u_i-1)^b\,
\prod_{i<j}(u_i-u_j)^{2\rho}
\times\cr
&\times
\prod_{i=1}^{p-1}\,t_i^a\,(1-t_i)^b
\prod_{i<j}(t_i-t_j)^{2\rho}\,.
\cr}
\eqn\cincuenta
$$
Notice that in \cincuenta\ the coefficients multiplying the
leading power of $z$ factorize in the variables $\{u_i\}$ and 
$\{t_i\}$. 

In order to get the values of $N_p$ and $\gamma_p$,
let us study in detail the $z\rightarrow 0$ expansion of the
function $\eta_p(\{u_i\},\{zt_i\})$. The key point in this
analysis is the fact that the rescaling $v_i=zt_i$ introduces
in $\eta_p$ a $z$-dependence, whose leading term we want to
determine. Due to the presence of the fermionic correlator in
\ccuatro, one must distinguish two cases, depending on the even
or odd character of the number $p-1$ of the $v_i$ variables.
Let us strat with the case in which $p-1$ is even. The
first relevant observation we must make is that, when the field
$w(zt_i)$ is contracted with $\chi(0)$, a $z^{-1}$ factor is
generated. Therefore, in the leading term, all the $w(v_i)$
fields must be contracted to $\chi(0)$. Moreover, it is clear
that the dominant power of $z$ coming from the fermionic
correlator is generated when the maximum number of $\psi(v_i)$
and $\bar\psi(v_i)$ are contracted among themselves. 

Our previous discussion shows that we have two sources of
powers of $z$ which, actually, are not independent. Indeed, it
is clear from \ccuatro\ that, for each $\psi$ field in the
fermionic correlator, we must have a $w$ field evaluated at the
same point. One might wonder if it is possible to have an
excess of $\psi(v_i)$'s with respect to the
$\bar\psi(v_i)$'s,  since the power of $z$ lost in the
fermionic correlator could be compensated by the contraction
of the extra $w(v_i)$'s to
$\chi(0)$. It turns out, however, that these terms, which by a
simple power counting could be present, do not contribute
because the corresponding fermionic correlators are zero at
leading order. Let us illustrate this point with an example.
Suppose that the number of $\psi(v_i)$'s is greater in two
units than those of the $\bar\psi(v_i)$'s. This means that two 
$\psi(v_i)$'s must be contracted with two  $\bar\psi(u_i)$'s.
The fermionic correlator of these contributions must contain
pieces of the type 
$<\,\bar\psi(u_k)\,\bar\psi(u_l)\,\psi(v_i)\,\psi(v_j)\,>$.
After substituting $v_i=zt_i$ and taking the limit
$z\rightarrow 0$ one gets:
$$
{\rm lim}_{z\rightarrow 0}\,
<\,\bar\psi(u_k)\,\bar\psi(u_l)\,
\psi(zt_i)\,\psi(zt_j)\,>\,=\,0\,,
\eqn\ciuno
$$
which can be regarded as a consequence of the antisymmetric
character of the fermionic correlators. It is
thus clear that only those permutations in \ccuatro\ having 
${p-1\over 2}$ $\psi(v_i)$ and $\bar\psi(v_i)$ fields in the
fermionic expectation value contribute to the leading term
when $z\rightarrow 0$ (recall that we are considering the case
in which $p-1$ is even).  It is not difficult to convince
oneself that, in the leading term, the fermionic correlator
factorizes into the product of two vacuum expectation values,
each of which involving fields that take values either in the 
$(1,\infty)$ or $(0,1)$ intervals of the real line. The
expression that one arrives at is:
$$
\eqalign{
&\eta_p(\{u_i\},\{zt_i\})\,\sim\,\,
(-1)^{{n\over 2}}\,\,
z^{1-p}\,
{(2j_1)!\over (2j_1-{p-1\over 2})!}\,\,
\times\cr\cr
&\times\{\,
<\,(\chi(0))^{2j_1-{p-1\over 2}}
\,(\chi(1))^{2j_2}\,
(w(\infty))^{2j_1+s}\,w(u_1)\,
\cdots\,w(u_{{n-p+1\over 2}})\,>
\times\cr\cr
&\times\,<\,\psi(u_1)\cdots\psi(u_{{n-p+1\over 2}})\,
\bar\psi(u_{{n-p+3\over 2}})\cdots\bar\psi(u_{n-p+1})\,>\,
 +\,{\rm permutations}\,\}\,
\times\cr\cr
&\times\,\{\,\prod_{i=1}^{{p-1\over 2}}\,t_i^{-1}\,\,
<\,\psi(t_1)\cdots\psi(t_{{p-1\over 2}})\,
\bar\psi(t_{{p+1\over 2}})\cdots\bar\psi(t_{p-1})\,>\,
+\,{\rm permutations}\,\}\,.
\cr\cr}
\eqn\cidos
$$

The origin of the different terms in eq. \cidos\ is clear. The
combinatorial and $t_i^{-1}$ factors come from the
contractions between $w(zt_i)$ and $\chi(0)$. The power of $z$
displayed in \cidos\ has a double origin: a factor 
$z^{{1-p\over 2}}$ comes from the contractions of 
the $w(zt_i)$'s  with $\chi(0)$,  whereas another 
$z^{{1-p\over 2}}$ contribution is generated in the fermionic
correlator. 

The situation when $p-1$ is odd is slightly different. In this
case, the number of fermionic operators taking values in the 
$(0,z)$ interval is odd. According to the general arguments
given above, the leading term is generated when the number of
$\psi$ fields in the $(0,z)$ segment exceeds in one unit to
that of the $\bar\psi$'s.  Notice that now the 
antisymmetry argument employed in \ciuno\ does not work.
Moreover, the fermionic correlator multiplying the leading term
does not factorize in a naive way. There is, however, a form of
writing a factorized expression for this quantity. It consists
in the introduction of two spectator fermions, located at the
points $z=0$ and $z=\infty$ of the complex plane, that are
inserted in each of the two correlators in which the initial
vacuum expectation value splits. In fact one can prove that,
for any pair of two positive integers 
$N$ and $M$, one has:
$$
\eqalign{
&<\,\psi(u_1)\cdots\psi(u_{N})\,
\bar\psi(u_{N+1})\cdots\bar\psi(u_{2N+1})
\psi(v_1)\cdots\psi(v_{M})\,
\bar\psi(v_{M+1})\cdots\bar\psi(v_{2M-1})\,>\,\sim\cr\cr
&\sim\,z^{1-M}\,
{\rm lim}_{R\rightarrow 0}\,
<\,\psi(R)\psi(u_1)\cdots\psi(u_{N})\,
\bar\psi(u_{N+1})\cdots\bar\psi(u_{2N+1})\,>\times\cr\cr
&\times{\rm lim}_{R\rightarrow \infty}\,R
<\,\psi(t_1)\cdots\psi(t_{M})\,
\bar\psi(t_{M+1})\cdots\bar\psi(t_{2M-1})\,\bar\psi(R)>\,,
\cr}
\eqn\citres
$$
where we have only kept the leading order term in $z$ and the
variables $v_i$ and $t_i$ are related as above (\ie\ 
$v_i=zt_i$ ). The proof of \citres\ is straightforward and can
be performed, for example, by means of the Cauchy determinant
formula (see eq. (B16)). We shall need \citres\ for
$N={n-p\over2}$ and $M={p\over 2}$. Thus, for $p-1$ odd, we
obtain:
$$
\eqalign{
&\eta_p(\{u_i\},\{zt_i\})\,\sim\,\,
(-1)^{{n\over 2}}\,\,z^{1-p}\,
{(2j_1)!\over (2j_1-{p\over 2})!}\,\,
\times\cr\cr
&\times\{\,
<\,(\chi(0))^{2j_1-{p\over 2}}\,(\chi(1))^{2j_2}\,
(w(\infty))^{2j_1+s}\,w(u_1)\,
\cdots\,w(u_{{n-p\over 2}})\,>
\times\cr\cr
&\times\,
{\rm lim}_{R\rightarrow 0}\,
<\,\psi(R)\psi(u_1)\cdots\psi(u_{{n-p\over 2}})\,
\bar\psi(u_{{n-p\over 2}+1})\cdots\bar\psi(u_{n-p+1})\,>\,
+\,{\rm permutations}\,\}
\times\cr\cr
&\times\,\{\,{\rm lim}_{R\rightarrow \infty}\,R\,
\prod_{i=1}^{{p\over 2}}\,t_i^{-1}\,\,
<\,\psi(t_1)\cdots\psi(t_{{p\over 2}})\,
\bar\psi(t_{{p\over 2}+1})\cdots\bar\psi(t_{p-1})\,
\bar\psi(R)>\,
+\,{\rm permutations}\,\}.
\cr\cr}
\eqn\cicuatro
$$
It is interesting to point out that the power of $z$ appearing
in \cidos\ and \cicuatro\ is the same, \ie\ 
$\eta_p\,\sim\,z^{1-p}$ for any value of $p$. Moreover, in the
change of variables $v_i\rightarrow zt_i$ a Jacobian factor is
introduced. Indeed one has:
$$
\int_0^{z}\,dv_1\cdots\int_0^{v_{p-2}}\,dv_{p-1}\,
[\,\cdots\,]\,
=\,
z^{p-1}\,
\int_0^{1}\,dt_1\cdots\int_0^{t_{p-2}}\,dt_{p-1}\,
[\,\cdots\,]\,.
\eqn\cicinco
$$
The power of $z$ in eq. \cicinco\ just cancels the one in 
$\eta_p$. Therefore, $\gamma_p$ can be read from the
expression of the leading term of $\lambda_p$ (eq.
\cincuenta):
$$
\gamma_p\,=\,8j_1j_2\rho\,+\,(p-1)\,
[\,a\,+\,b\,+\,(p-2)\rho\,]\,.
\eqn\ciseis
$$
After some elementary algebraic manipulations, the exponents
$\gamma_p$ can be written as differences of 
${\rm osp}(1\vert 2)$ conformal weights:
$$
\gamma_p\,=\,\Delta_{j_3}\,-\,
\Delta_{j_1}\,-\,\Delta_{j_2}\,,
\eqn\cisiete
$$
where the quantities $\Delta_{j_i}$ are given in \ocho\ and
$j_3$, as a function of $p$, is:
$$
j_3\,=\,j_1\,+\,j_2\,+{1\,-\,p\over 2}\,.
\eqn\ciocho
$$
Eq. \cisiete\ allows to interpret $j_3$ as the isospin of the
s-channel intermediate state. Notice that as $p=1,\cdots,
4j_2+1$ the values taken by $j_3$ are  
$j_1\,-\,j_2,\,j_1\,-\,j_2\,+{1\over 2},\cdots,\,j_1+j_2$.
Remarkably, these are the values of the isospin that appear
in the Clebsch-Gordan decomposition of the tensor product of
two ${\rm osp}(1\vert 2)$ representations of isospins $j_1$
and $j_2$. This result confirms our analysis of the
three-point functions and encourages us  to proceed with the
study of the four-point function.

The coefficients $N_p$ will be given by  multiple integrals,
whose explicit expressions can be obtained by gathering the
different contributions coming from our previous equations.
Let us consider first the case in which $p-1$ is even. 	In
general, $N_p$ will depend on the product of two integrals: one
over the variables $u_i$ and the other involving the $t_i$'s.
The expression of the latter can be obtained by collecting the
factors depending on $t_i$ of eqs. \cincuenta\ and \cidos. The
result is:
$$
\eqalign{
&S_p\,\equiv\,\int_0^{1}\,dt_1\cdots
\int_0^{t_{p-2}}\,dt_{p-1}\,\times\cr
&\times\Bigl\{\,\prod_{i=1}^{{p-1\over 2}}\,t_i^{a-1}\,
\prod_{i={p-1\over 2}+1 }^{p-1}\,t_i^{a}\,
<\,\psi(t_1)\cdots\psi(t_{{p-1\over 2}})\,
\bar\psi(t_{{p+1\over 2}})\cdots\bar\psi(t_{p-1})\,>\,
+\,{\rm permutations}\Bigr\}\times\cr
&\times
\prod_{i=1}^{p-1}\,(1-t_i)^b
\prod_{i<j}(t_i-t_j)^{2\rho}\,.
\cr}
\eqn\cinueve
$$
$S_p$ is given by an integral of the type studied by
Selberg
\REF\sel{A. Selberg\journal\nmt&26(44)71.}
[\sel]. In  appendix B we have studied the integrals of
this kind that we shall need in our calculation. Actually, 
$S_p$ belongs to the family of even-dimensional integrals
$J_{2r}^m$ defined in eq. (B11). Indeed, comparing the
right-hand sides of eqs. \cinueve\ and (B11), one finds:
$$
S_p\,=\,J_{p-1}^{{p-1\over 2}}\,(a-1,b,\rho)\,.
\eqn\sesenta
$$
In  appendix B we have determined the value of the
functions $J_{2r}^m$ in terms of the Euler $\Gamma$-function
(see eq. (B12)).

In order to find the contribution of the integral
over the $(1,\infty)$ interval to $N_p$, one has to evaluate
explicitly the expectation value of the $w$ and $\chi$ fields
appearing in eqs. \cidos\ and \cicuatro. Apart from an
irrelevant constant, this quantity can be written as:
$$
\eqalign{
&<\,(\chi(0))^{2j_1-[{p\over 2}]}\,(\chi(1))^{2j_2}\,
(w(\infty))^{2j_1+s}\,w(u_1)\,
\cdots\,w(u_{[{n-p+1\over 2}]})\,>\,=\,\cr\cr
&=\,\sum_{m=0}^{[{n-p+1\over 2}]}\,B_m\,\Bigl\{\,
\prod_{i=1}^{m}\,u_i^{-1}\,
\prod_{i=m+1}^{[{n-p+1\over 2}] }\,(u_i-1)^{-1}\,+\,
{\rm permutations}\,\Bigr\}\,,
\cr}
\eqn\suno
$$
where the combinatorial constants $B_m$ are given by:
$$
B_m\,\,=\,\,
{(2j_1-[{p\over 2}])!\over
(2j_1\,-\,[{p\over 2}]-m)!}\,\,\,
{(2j_2)!\over
(2j_2\,-\,[{n-p+1\over 2}]+m)!}\,.
\eqn\sdos
$$
Notice that \suno\ is the $w\chi$ correlator needed for an
arbitrary value of $p$ (\ie\ eq. \sdos\ can be used both in
\cidos\ and
\cicuatro). Coming back to the case in which $p-1$ is even,
after eq. \suno\ is substituted in \cidos, one realizes that
the integrals to be computed are:
$$
\eqalign{
&{\cal S}_p^m\,\equiv\,
\int_1^{\infty}\,du_1\cdots
\int_1^{u_{n-p}}\,du_{n-p+1}\,\,
\Bigl\{\,\prod_{i=1}^{n-p+1}\,
u_i^{a+b+2\rho(p-1)}\,(u_i-1)^b\,\,\times\cr
&\times\,[\,
\prod_{i=1}^{m}\,u_i^{-1}\,
\prod_{i=m+1}^{{n-p+1\over 2} }\,(u_i-1)^{-1}\,+\,
{\rm permutations}\,]\,\times\cr
&\times\,
<\,\psi(u_1)\cdots\psi(u_{{n-p+1\over 2}})\,
\bar\psi(u_{{n-p+3\over 2}})\cdots\bar\psi(u_{n-p+1})\,>\,
 +\,{\rm permutations}\,\Bigr\}\,\times\cr
&\times\,\prod_{i<j}(u_i-u_j)^{2\rho}\,.
\cr}
\eqn\stres
$$
We would like to recast ${\cal S}_p^m$ as a multiple ordered
integral in the $(0,1)$ interval. This can be achieved after
a change of variables that involves an inversion and
reordering of the $u_i$'s. Let us introduce new variables
$\xi_i$, $i=1,\cdots, n-p+1$, by means of the equation:
$$
\xi_i\,=\,[\,u_{n-p+2-i}\,]^{-1}\,.
\eqn\scuatro
$$
Notice that, in the integration domain appearing in the
definition of ${\cal S}_p^m$, the variables $\xi_i$ satisfy the
inequalities $1\ge\xi_1\ge\cdots\ge\xi_{n-p+1}\ge 0$. After an
straightforward calculation one can rewrite eq. \stres\ as:
$$
\eqalign{
&{\cal S}_p^m\,=\,
\int_0^{1}\,d\xi_1\cdots
\int_0^{\xi_{n-p}}\,d\xi_{n-p+1}\,\Bigl\{\,
\prod_{i=1}^{{n-p+1\over 2}}\,\xi_i^{\bar a}\,\,
\xi_{{n-p+1\over 2}+i}^{\bar a+1}\,\,(1-\xi_i)^b
\times\cr
&\times\,[\,
\prod_{i=1}^{m}\,(\,1-\xi_{{n-p+1\over 2}+i}\,)^{b}
\prod_{i=m+1}^{{n-p+1\over 2} }\,
(\,1-\xi_{{n-p+1\over 2}+i}\,)^{b-1}
\,+\,{\rm permutations}\,]\,\times\cr
&\times\,
<\,\psi(\xi_1)\cdots\psi(\xi_{{n-p+1\over 2}})\,
\bar\psi(\xi_{{n-p+3\over 2}})\cdots\bar\psi(\xi_{n-p+1})\,>\,
 +\,{\rm permutations}\,\Bigr\}\,\times\cr
&\times\,\prod_{i<j}(\xi_i-\xi_j)^{2\rho}\,,
\cr}
\eqn\scinco
$$
where $\bar a$ is defined as:
$$
\bar a\,=\,-1-a-2b-2\rho(n-1)\,.
\eqn\sseis
$$
Eq. \scinco\ allows to identify ${\cal S}_p^m$ with an
integral of those computed in appendix B. In fact, one has:
$$
{\cal S}_p^m\,=\,{\cal J}_{n-p+1}^{m}\,(\bar a,b,\rho)\,.
\eqn\ssiete
$$
The functions ${\cal J}_{2r}^{m}\,( a,b,\rho)$ have been
defined in eq. (B36). Their values, again given in terms of
$\Gamma$-functions, have been written down in eq. (B37).

For $p-1$ odd one can proceed similarly. The odd-dimensional
integrals needed to compute $N_p$ match  the definitions
of the functions $J_{2r+1}^m$ and ${\cal J}_{2r+1}^m$ adopted
in appendix B (eqs. (B41) and (B43) respectively). The only
subtle point in the calculation of $N_p$ appears when one
studies the behaviour of the fermionic correlator under the
change of variables \scuatro. Actually, one can prove an
equation relating the correlator in the variables $u_i$, 
with an additional fermionic spectator at the origin,  to a
correlator of the fields $\psi(\xi_i)$ and
$\bar\psi(\xi_i)$, which has  a fermionic insertion at
infinity. This equation is:
$$
\eqalign{
&{\rm lim}_{R\rightarrow 0}\,
<\,\psi(R)\psi(u_1)\cdots\psi(u_{{n-p\over 2}})\,
\bar\psi(u_{{n-p\over 2}+1})\cdots\bar\psi(u_{n-p+1})\,>\,
=\,\cr\cr
&=\,\Bigl(\,\prod_{i=1}^{n-p+1}\xi_i\Bigr)\,\,
{\rm lim}_{R\rightarrow \infty}\,R
<\,\psi(\xi_1)\cdots\psi(\xi_{{n-p\over 2}+1})\,
\bar\psi(\xi_{{n-p\over 2}+2})
\cdots\bar\psi(\xi_{n-p+1})\,
\bar\psi(R)>\,.\cr}
\eqn\socho
$$
It is a highly non-trivial fact (see appendix B) that there
exist closed expressions for the functions $J_N^m$ and 
${\cal J}_N^m$ for even and odd values  of 
$N$. These expressions are written in eqs. (B42) and (B44).
Moreover, one can put $N_p$ for arbitrary $p$ in terms of these
functions:
$$
N_p\,=\,(-1)^{{n\over 2}}\,\,
{(2j_1)!\over
(2j_1\,-\,[{p\over 2}])!}\,\,\,
J_{p-1}^{[{p-1\over 2}]}\,(a-1,b,\rho)\,\,\,
\sum_{m=0}^{{[{n-p+1\over 2}]}}\,B_m\,\,
{\cal J}_{n-p+1}^{m}\,(\bar a,b,\rho)\,.
\eqn\snueve
$$
In the next section we shall use the value of $N_p$ written in
eq. \snueve\ to obtain the operator algebra of the model.

\chapter{The operator product algebra}

So far in our study of the ${\rm osp}(1\vert 2)$ correlators
we have only considered the holomorphic sector of the theory. 
In order to compute the physical correlation functions
$G(z\,,\,\bar z)$, one must combine   the holomorphic and
antiholomorphic blocks in such a way that $G(z\,,\,\bar z)$
becomes monodromy invariant.
We shall restrict ourselves to operators whose $z$ and 
$\bar z$ quantum numbers are the same. This, in particular,
means that they will have equal holomorphic and antiholomorphic
conformal dimensions, \ie\ the operators we shall consider
will have zero conformal spin. 
According to the general
arguments valid for the Coulomb gas representations, one
constructs $G(z\,,\,\bar z)$ as a quadratic expression of the
functions $I_p(z)$ and their complex conjugates:
$$
G(z\,,\,\bar z)\,=\,\sum_p\,X_p\,|\,I_p(z)\,|^2\,,
\eqn\setenta
$$
where the coefficients $X_p$ are such that the right-hand
side of eq. \setenta\ is monodromy invariant. In ref. [\DF] a
technique to compute these coefficients  $X_p$  has been
developed. This method is based on the representation of the
functions $I_p(z)$ as contour integrals in the complex plane, 
and allows to obtain the $X_p$'s up to a global (\ie\
$p$-independent) constant. It is straightforward to adapt the
results of [\DF] to our case. One gets:
$$
\eqalign{
X_p\,=&\,\prod_{i=1}^{p-1}\,s(i(\rho-{1\over 2})\,)\,
\prod_{i=0}^{p-2}\,
{s(a+i(\rho-{1\over 2})\,)\,s(1+b+i(\rho-{1\over 2})\,)\over
s(1+a+b+(p-2+i)(\rho-{1\over 2})\,)}\,\times\cr\cr
\times&\,\prod_{i=1}^{n-p+1}\,s(i(\rho-{1\over 2})\,)\,
\prod_{i=0}^{n-p}\,
{s(1-a-2b-(\rho-{1\over 2})(2(n-1)-i)\,)\,
s(b+i(\rho-{1\over 2})\,)\over
s(1-a-b-(\rho-{1\over 2})(2(p-1)+i)\,)}\,,\cr}
\eqn\stuno
$$
where we have introduced the notation
$$
s(x)\,\equiv\,sin\,(\pi x)\,.
\eqn\stdos
$$
It is now easy to get the leading $|z|\rightarrow
0$ contributions to $G(z\,,\,\bar z)$  of the different
channels. Indeed, after combining eqs. \cnueve, \cisiete\ and
\setenta, one obtains:
$$
G(\,z\,,\,\bar z\,)\,\sim\,\sum_p\,\,\Bigl[\,
{S_p\over |z|^{2\,(\Delta_{j_1}+\Delta_{j_2}
-\Delta_{j_3})}}\,+\,O(z)\,\Bigr]\,.
\eqn\sttres
$$
In eq. \sttres\ $j_3$ and $p$ are related as in eq. \ciocho\
and the coefficients $S_p$ are:
$$
S_p\,=\,X_p\,(N_p)^2\,.
\eqn\stcuatro
$$
The operator product algebra of the model can be determined
by comparing the expansion of eq. \sttres\ with the one
obtained by performing some appropriate OPE's of primary
fields inside the correlator $G(\,z\,,\,\bar z\,)$. In
general, the operator product algebra of the theory will have
the form:
$$
\Phi_{m_1}^{j_1}(z_1,\bar z_1)\,\Phi_{m_2}^{j_2}(z_2,\bar z_2)\,=\,
\sum_{j_3,m_3}\,D_{j_1,m_1;j_2,m_2}^{j_3,m_3}\,\,\Bigl[
\,{\Phi_{m_3}^{j_3}(z_2,\bar z_2)\over 
|z_1\,-z_2|^{2(\Delta_{j_1}\,+\,\Delta_{j_2}\,-\,\Delta_{j_3})}}\,+\,
O(z_1\,-z_2)\,\Bigr].
\eqn\stcinco
$$
The coefficients $D_{j_1,m_1;j_2,m_2}^{j_3,m_3}$ in eq.
\stcinco\ are the so-called structure constants of the model.
Their determination from the quantities $S_p$ is the main
objective of the present section. The value of the structure
constants depends on the normalization chosen for the
two-point correlator 
$<\,\Phi_{m_1}^{j_1}(z_1,\bar z_1)\,
\Phi_{m_2}^{j_2}(z_2,\bar z_2)\,>$. The standard choice for
the normalization of these functions is:
$$
<\,\Phi_{m_1}^{j_1}(z_1,\bar z_1)\,
\Phi_{m_2}^{j_2}(z_2,\bar z_2)\,>
\,={\delta_{j_1,j_2}\,\delta_{m_1,-m_2}\over
|z_1\,-z_2|^{4\Delta_{j_1}}}\,,
\eqn\stseis
$$
which implies the following constraint for the structure
constants:
$$
D_{j_1,m_1;j_1,-m_1}^{0,0}\,=\,1\,.
\eqn\stsiete
$$
For primary fields $\Phi_{m}^{j}$ that correspond to states
$|j,m>$ with negative norm we shall include a minus sign in
the right-hand side of \stseis. As  is explained in  
appendix A, these negative norm states appear in the odd
representations of ${\rm osp}(1\vert 2)$. Although the
representations involved in our correlator 
$G(\,z\,,\,\bar z\,)$ are even, odd representations do appear
in the intermediate states $|j_3,m_3>$ and, therefore, we must
be careful with this sign (see below). On the other hand, we
can reobtain the power behaviour \sttres\ by substituting the
OPE's 
$\Phi_{-j_1}^{j_1}(z_1,\bar z_1)
\,\Phi_{j_2}^{j_2}(z_2,\bar z_2)$ and
$\Phi_{-j_2}^{j_2}(z_3,\bar z_3)
\,\Phi_{j_1}^{j_1}(z_4,\bar z_4)$ for $z_1=0$, $z_2=z$, 
$z_3=1$ and $z_4=\infty$ in the correlator 
$G(\,z\,,\,\bar z\,)$. The equation we arrive at is: 
$$
G(\,z\,,\,\bar z\,)\,\sim\,\sum_{j_3,m_3}\,\,
(-1)^{\sigma(j_3,m_3)}\,\,\Bigl[\,
{[\,D_{j_1,j_1;j_2,-j_2}^{j_3,m_3}\,]^2\over
|z|^{2\,(\Delta_{j_1}+\Delta_{j_2}
-\Delta_{j_3})}}+\,O(z)\,\Bigr]\,,
\eqn\stocho
$$
where $\sigma(j_3,m_3)$ is $0$($1$) if the state $|j_3,m_3>$
has positive(negative) norm. It is clear by comparing eqs.
\sttres\ and \stocho\ that one has the identification:
$$
(-1)^{\sigma(j_3,m_3)}\,\,
[\,D_{j_1,j_1;j_2,-j_2}^{j_3,m_3}\,]^2\,\,\sim
\,\,S_p\,,
\eqn\stnueve
$$
which allows to get the structure constants in terms of the
quantities $S_p$. Notice, however, that the global
factor ambiguity in the determination of the $X_p$'s  is
inherited in the constants $S_p$. As  will be discussed
below, this ambiguity can be eliminated by imposing the
normalization condition \stsiete.

Before plunging into the calculation of the coefficients $S_p$
and the structure constants, let us obtain a remarkable
simplification of the expression of the $N_p$'s given in
\snueve. First of all, let us rewrite eq. \snueve\ using the
explicit expression of the coefficients $B_m$ written in eq.
\sdos:
$$
\eqalign{
N_p\,=\,(-1)^{{n\over 2}}\,\,
&J_{p-1}^{[{p-1\over 2}]}\,(a-1,b,\rho)\,\,\times\cr\cr
&\times\,\,\sum_{m=0}^{{[{n-p+1\over 2}]}}
{(2j_1)!\over
(2j_1\,-\,[{p\over 2}]-m)!}\,\,\,
{(2j_2)!\over
(2j_2\,-\,[{n-p+1\over 2}]+m)!}\,\,\,
{\cal J}_{n-p+1}^{m}\,(\bar a,b,\rho)\,.
\cr}
\eqn\ochenta
$$
The value of the integrals 
${\cal J}_{n-p+1}^{m}\,(\bar a,b,\rho)$ is given in eq. (B44).
From this equation it is easy to relate these integrals for
arbitrary $m$ to the same functions with $m=0$. Indeed, using
in (B44) the elementary properties of the $\Gamma$-function,
one gets:
$$
{\cal J}_{n-p+1}^{m}\,(\bar a,b,\rho)\,=\,
{[{n-p+1\over 2}]\choose m}\,\,\,
\prod_{i=0}^{[{n-p+1\over 2}]-m-1}\,\,\,
{1+\bar a+b+2\rho([{n-p+1\over 2}]+i)\over
b+2\rho i}\,\,\,
{\cal J}_{n-p+1}^{0}\,(\bar a,b,\rho)\,.
\eqn\ouno
$$
Employing eq. \ouno\ we will be able to perform the summation
in $m$ in eq. \ochenta. Our first step will consist in
writing the product \ouno\ in terms of factorials. Using the
expressions of $\bar a$(eq. \sseis), $b$(eq. \cocho) and
$\rho$(eq. \csiete) in terms of $j_1$ and $j_2$ one gets:
$$
\eqalign{
\prod_{i=0}^{[{n-p+1\over 2}]-m-1}\,\,\,
&{1+\bar a+b+2\rho([{n-p+1\over 2}]+i)\over
b+2\rho i}\,\,\,=\cr\cr
&=\,\,(-1)^{[{n-p+1\over 2}]-m}\,\,
{(2j_1-[{p\over 2}]+[{n-p+1\over 2}]-m)!\over
(2j_1\,-\,[{p\over 2}])!}\,\,\,
{(2j_2-[{n-p+1\over 2}]+m)!\over
(2j_2)!}\,\,.\cr}
\eqn\odos
$$
Substituting the right-hand side of eq. \odos\ in eq.
\ochenta\ one realizes that the factors depending on $j_2$
disappear. Moreover, the sum to be computed can be written as:
$$
\eqalign{
&\sum_{m=0}^{{[{n-p+1\over 2}]}}\,\,
(-1)^m\,{[{n-p+1\over 2}]\choose m}\,\,\,
{(2j_1-[{p\over 2}]+[{n-p+1\over 2}]-m)!\over
(2j_1\,-\,[{p\over 2}]-m)!}\,=\,\cr\cr
&=\,\sum_{m=0}^{{[{n-p+1\over 2}]}}\,\,
{[{n-p+1\over 2}]\choose m}\,\,\,
\prod_{i=0}^{m-1}\,\,(-2j_1+[{p\over 2}]+i)\,\,\,
\prod_{i=0}^{[{n-p+1\over 2}]-m-1}
\,\,(2j_1-[{p\over 2}]+1+i)\,.
\cr}
\eqn\otres
$$
In order to evaluate the right-hand side of \otres,  we shall
use the identity (which can be easily demonstrated by
induction):
$$
\sum_{m=0}^{N}\,\,{N\choose m}\,
\prod_{i=0}^{m-1}\,(\,A\,+\,i\rho\,)\,
\prod_{i=0}^{N-m-1}\,(\,B\,+\,i\rho\,)\,\,=\,
\prod_{i=0}^{N-1}\,(\,A\,+\,B\,+\,i\rho\,)\,.
\eqn\ocuatro
$$
Putting  $A=-B+1=-2j_1+[{p\over 2}]$,
$N=[{n-p+1\over 2}]$ and $\rho=1$ in \ocuatro, one can get 
the value of the sum \otres, namely:
$$
\sum_{m=0}^{{[{n-p+1\over 2}]}}\,\,
(-1)^m\,{[{n-p+1\over 2}]\choose m}\,\,\,
{(2j_1-[{p\over 2}]+[{n-p+1\over 2}]-m)!\over
(2j_1\,-\,[{p\over 2}]-m)!}\,=\,\,
\Bigl([{n-p+1\over 2}]\Bigr)!\,.
\eqn\ocinco
$$
Using this result in \ochenta, and taking into account that
${\cal J}_{n-p+1}^{0}\,(\bar a,b,\rho)\,=\,
J_{n-p+1}^{[{n-p+1\over 2}]}\,(\bar a,b,\rho)$ (compare the
definitions of both integrals in  appendix B), one can
obtain the simplified expression of $N_p$ we were looking
for, \ie:
$$
N_p\,=\,(-1)^{[{p\over 2}]}\,\,
{(2j_1)!\,({n\over 2}\,-\,[{p\over 2}])!\over
(2j_1\,-\,[{p\over 2}])!}\,\,\,
J_{p-1}^{[{p-1\over 2}]}\,(a-1,b,\rho)\,\,\,
J_{n-p+1}^{[{n-p+1\over 2}]}\,(\bar a,b,\rho)\,\,.
\eqn\oseis
$$
We can now use the value of the integrals $J_N^m$, which has
been written in eq. (B42), to get the following
representation of $N_p$:
$$
\eqalign{
N_p\,=&\,(-1)^{[{p\over 2}]}\,\,
{(2j_1)!\,({n\over 2}\,-\,[{p\over 2}])!\over
(2j_1\,-\,[{p\over 2}])!}\,\,\,
\mu_{p-1}(\rho)\,\,\mu_{n-p+1}(\rho)\,\times\cr
&\times\prod_{i=0}^{p-2}\,
{\Gamma(a+i(\rho-{1\over 2})+[{i+1\over 2}])\,
\Gamma(1+b+i(\rho-{1\over 2})+[{i\over 2}])\over
\Gamma(a+b+(\rho-{1\over 2})(p-2+i)+[{i+p\over 2}])}\,\times\cr
&\times\prod_{i=0}^{n-p}\,
{\Gamma(1+\bar a+i(\rho-{1\over 2})+[{i+1\over 2}])\,
\Gamma(1+b+i(\rho-{1\over 2})+[{i\over 2}])\over
\Gamma(1+\bar a+b+(\rho-{1\over 2})(n-p+i)
+[{i+n-p+2\over 2}])}\,\,.
\cr}
\eqn\osiete
$$
The function $\mu_{N}(\rho)$ appearing in eq. \osiete\ has
been defined in eq. (B13). Having the expression \osiete\ at
our disposal, we can resume our calculation of the
constants $S_p$. We must substitute  $X_p$ and $N_p$, as given
in eqs. \stuno\ and \osiete, in the right-hand side of eq.
\stcuatro. In this calculation we shall make use of the
relation 
$s(x)\,[\,\Gamma (x)\,]^2\,=
\,\pi\,\Gamma(x)\,/\,\Gamma(1-x)$. The final result can be
compactly written in terms of the functions:
$$
\eqalign{
\Pi_{N}(a,b,\rho)\,\equiv\,&
 \prod_{i=0}^{N}\,
{\Gamma(1+ a+i(\rho-{1\over 2})+[{i+1\over 2}])\,\over
\Gamma(- a-i(\rho-{1\over 2})-[{i+1\over 2}])}\,\,
{\Gamma(1+b+i(\rho-{1\over 2})+[{i\over 2}])\,\over
\Gamma(-b-i(\rho-{1\over 2})-[{i\over 2}])}\,\times\cr
&\times\prod_{i=0}^{N}\,
{\Gamma(- a-b-(\rho-{1\over 2})(N+i)-[{i+N+2\over 2}])
\,\over
\Gamma(1+ a+b+(\rho-{1\over 2})(N+i)+[{i+N+2\over 2}])}\cr\cr
\widehat\mu_N(\rho)\,\equiv\,&
\,\prod_{i=1}^{N}\,
{\Gamma(i(\rho+{1\over 2})-[{i\over 2}])\,\over
\Gamma(1-i(\rho+{1\over 2})+[{i\over 2}])}\,\,.
\cr}
\eqn\oocho
$$
With this definition, the constants $S_p$ are given by:
$$
\Bigl({\Gamma(\rho+{1\over 2})\over \pi}\Bigr)^{2n}\,S_p\,=\,
\Bigl(\,{(2j_1)!\,({n\over 2}\,-\,[{p\over 2}])!\over
(2j_1\,-\,[{p\over 2}])!\,}\Bigr)^{2}\,\,\,
\widehat\mu_{p-1}(\rho)\,\,\widehat\mu_{n-p+1}(\rho)\,
\,\,\Pi_{p-2}(a-1,b,\rho)\,\,
\Pi_{n-p}(\bar a,b,\rho)\,,
\eqn\onueve
$$
where we have multiplied $S_p$ by a $p$-independent constant,
which  disappears when these quantities are properly
normalized. We shall need the values of $S_p$ in terms of
$j_1$, $j_2$ and $j_3$. Substituting the values of $a$, $\bar
a$, $b$, $p$ and $\rho$ as functions of the isospins and
recalling that $n=4j_2$, one can rewrite eq. \onueve\ as:
$$
\eqalign{
&\Bigl({\Gamma (\rho+{1\over 2})\over \pi}\Bigr)^{2n}\,
S(j_1,j_2;j_2,j_1|j_3)\,=\cr\cr
&=\,\Bigl({(2j_1)!\,([j_2+j_3-j_1])!\over
([j_1+j_3-j_2])!}\Bigr)^{2}\,\,
\widehat\mu_{2j_1+2j_2-2j_3}(\rho)\,
\widehat\mu_{2j_2+2j_3-2j_1}(\rho)\,
\times\cr\cr
&\times 
\prod_{i=0}^{2j_1+2j_2-2j_3-1}\,\Bigl[\,\,
{\Gamma(\rho(i-4j_1)-{i\over 2}+[{i+1\over 2}])\,\over
 \Gamma(1-\rho(i-4j_1)+{i\over 2}-[{i+1\over 2}])}\,\,
{\Gamma(1+\rho(i-4j_2)-{i\over 2}+[{i\over 2}])\,\over
 \Gamma(-\rho(i-4j_2)+{i\over 2}-[{i\over 2}])}\times\cr\cr
&\times{\Gamma(\rho(4j_3+2+i)-{i\over 2}+[{i+1\over 2}])\,\over
 \Gamma(1-\rho(4j_3+2+i)+{i\over 2}-[{i+1\over 2}])}
\,\,\Bigr]\,\times\cr\cr
&\times\prod_{i=0}^{2j_2+2j_3-2j_1-1}\,\Bigl[\,\,
{\Gamma( \rho(4j_1+2+i)-{i\over 2}+[{i+1\over 2}])\,\over
 \Gamma(1-\rho(4j_1+2+i)+{i\over 2}-[{i+1\over 2}])}\,\,
{\Gamma(1+\rho(i-4j_2)-{i\over 2}+[{i\over 2}])\,\over
 \Gamma(-\rho(i-4j_2)+{i\over 2}-[{i\over 2}])}\times\cr\cr
&\times{\Gamma(\rho(i-4j_3)-{i\over 2}+[{i+1\over 2}])\,\over
 \Gamma(1-\rho(i-4j_3)+{i\over 2}-[{i+1\over 2}])}\,\,\Bigr]
\,,\cr}
\eqn\noventa
$$
where we have called $S(j_1,j_2;j_2,j_1|j_3)$ to what we were
previously denoting by $S_p$. 

Let us now consider  the
question of the normalization of the $S_p$'s needed to
convert the identification \stnueve\ in a true equality. As we
have previously mentioned, this normalization can be fixed by
requiring the fulfillment of eq. \stsiete. Following the
analysis of refs. [\DF, \Dotsenko], we shall achieve this by 
dividing the constants \noventa\ by $S(j_2,j_2;j_2,j_2|0)$,
which correspond to the conformal blocks with $j_1=j_2$ and the
trivial representation, \ie\ that with isospin $j_3=0$, in
the intermediate state. Moreover, let us notice that the 
state exchanged in the s-channel is $|j_3,j_1-j_2>$. In order
to fix the sign  in the left-hand side of eq. \stnueve, we
must determine under which conditions $|j_3,j_1-j_2>$ has
negative norm. This
state appears in the tensor product of $|j_1,j_1>$ and 
 $|j_2,-j_2>$, which are both bosonic since we are assuming
that they belong to even representations of 
${\rm osp}(1\vert 2)$. Thus the state $|j_3,j_1-j_2>$ is
bosonic. Recall that, in general, $|j,m>$ is bosonic when $j-m$
is integer (half-integer) for an even (odd) representation.
Moreover, as  is explained in  appendix A, one can
arrange the normalization conventions in such a way that only
those states  $|j,m>$ belonging to an odd representation and
with $j-m$ half-integer have negative norm. These results
imply in our case that  $|j_3,j_1-j_2>$ has negative norm
only when $j_3-j_1+j_2$ is half-integer and, therefore, in eq.
\stnueve\ we must take:
$$
(-1)^{\sigma(j_3,m_3)}\,\,=\,\,(-1)^{2j_1+2j_2-2j_3}\,\,.
\eqn\nuno
$$
All these considerations lead us to write:
$$
\Bigl[\,D_{j_1,j_1;j_2,-j_2}^{j_3,j_1-j_2}\,\Bigr]^2\,=\,
(-1)^{2j_1+2j_2-2j_3}\,\,
{S(\,j_1,j_2;j_2,j_1|j_3\,)
\over S(\,j_2,j_2;j_2,j_2|0\,)}\,\,.
\eqn\ndos
$$
In order to compute the right-hand side of eq. \ndos, it is
convenient to introduce the quantities $c_j$, defined as:
$$
c_j\,\equiv\,
\prod_{i=1}^{4j}\,\,
{\Gamma(1-{i\over 2}+i\rho+[{i\over 2}])\,\,
\Gamma({1+i\over 2}-(i+1)\rho-[{i\over 2}])\over
\Gamma({i\over 2}-i\rho-[{i\over 2}])\,\,
\Gamma({1-i\over 2}+(i+1)\rho+[{i\over 2}])}\,\,.
\eqn\ntres
$$
After some rearrangements of the products in \noventa, one
can put the normalization factor in \ndos\ in the form:
$$
\Bigl({\Gamma(\rho+{1\over 2})\over \pi}\Bigr)^{2n}\,\,
S(\,j_2,j_2;j_2,j_2|0\,)\,=\,{((2j_2)!)^2\over c_{j_2}}\,\,.
\eqn\ncuatro
$$
One can perform similar manipulations to the numerator of the
right-hand side of eq. \ndos. In fact, it is possible to
arrive at an expression in which most of the products appear
as a square power. This expression is:
$$
\eqalign{
&{S(\,j_1,j_2;j_2,j_1|j_3\,)
\over S(\,j_2,j_2;j_2,j_2|0\,)}\,=\,
\Bigl(\,{(2j_1)!\,([j_2+j_3-j_1])!\over
([j_1+j_3-j_2])!\,(2j_2)!}\,\Bigr)^{2}\,\,
{c_{j_1}\,c_{j_2}\over c_{j_3}}\,\,
\Bigl[\,\prod_{i=1}^{2j_1+2j_2-2j_3}\,
\!\!\!\!{\Gamma(i(\rho+{1\over 2})-[{i\over 2}])\,\over
\Gamma(1-i(\rho+{1\over 2})+[{i\over 2}])}\times\cr\cr
&\,\,\,\,\,\,\,\,\,\,\,\,\,\,\,\,\,\,\times
\prod_{i=0}^{2j_1+2j_2-2j_3-1}\,
{\Gamma(\rho(i-4j_1)-{i\over 2}+[{i+1\over 2}])\,\over
 \Gamma(1-\rho(i-4j_1)+{i\over 2}-[{i+1\over 2}])}\,\,
{\Gamma(1+\rho(i-4j_2)-{i\over 2}+[{i\over 2}])\,\over
 \Gamma(-\rho(i-4j_2)+{i\over 2}-[{i\over 2}])}\times\cr\cr
&\,\,\,\,\,\,\,\,\,\,\,\,\,\,\,\,\,\,
\times{\Gamma(\rho(4j_3+2+i)-{i\over
2}+[{i+1\over 2}])\,\over
 \Gamma(1-\rho(4j_3+2+i)+{i\over 2}-[{i+1\over 2}])}\,\,
\Bigr]^2\,\,.\cr}
\eqn\ncinco
$$
We would like to convert \ncinco\ into an equation symmetric in
the three isospins $j_1$, $j_2$ and $j_3$. In order to attain
this purpose let us introduce the functions $\lambda(j)$ and 
${\cal P}(j)$. The former is defined as:
$$
\lambda(j)\,\equiv\,
{\Gamma({j\over 2}\,+\,j\rho\,-\,[{j\over 2}])\over
\Gamma({j\over 2}\,-\,j\rho\,-\,[{j\over 2}])}\,\,,
\eqn\nseis
$$
while ${\cal P}(j)$ is given by:
$$
{\cal P}(j)\,\equiv\,\prod_{i=1}^{j}\,\lambda(i)\,=\,
\prod_{i=1}^{j}\,
{\Gamma({i\over 2}\,+\,i\rho\,-\,[{i\over 2}])\over
\Gamma({i\over 2}\,-\,i\rho\,-\,[{i\over 2}])}\,\,.
\eqn\nsiete
$$
It turns out that the different contributions in \ncinco\ can
be written in terms of these two functions and some
combinatorial factors. For example, the ratio of $c_j$'s
displayed in \ncinco\ can be put as:
$$
{c_{j_1}\,c_{j_2}\over c_{j_3}}\,\,=\,\,
(-1)^{2j_1+2j_2-2j_3}\,\,(2\rho)^{4j_1+4j_2-4j_3}\,\,
\Bigl({\,(2j_1)!\,(2j_2)!\over(2j_3)!\,}\Bigr)^2
\,\,\,
\lambda(1)\,\,
{\lambda(4j_3+1)\over \lambda(4j_1+1)\lambda(4j_2+1)}\,\,.
\eqn\nocho
$$
The total combinatorial factor multiplying the product of
$\lambda$ and ${\cal P}$ functions is:
$$
\Bigl[\,C_{j_1,j_1;j_2,-j_2}^{j_3,j_1-j_2}\,\Bigr]^2\,=
{\,(2j_1)!\,(2j_2)!\over
([j_1+j_2+j_3+{1\over 2}])!\,
([j_1+j_2-j_3])!}\,\,.
\eqn\nnueve
$$
Remarkably, the quantity $C_{j_1,j_1;j_2,-j_2}^{j_3,j_1-j_2}$
in eq. \nnueve\ is the ${\rm osp}(1\vert 2)$ Clebsch-Gordan
coefficient for the coupling of the three states 
$|j_1, j_1>$, $|j_2, -j_2>$ and $|j_3, j_1-j_2>$ (see
appendix A). With this identification, we can write the
structure constants for arbitrary values of the isospins and
Cartan components as:
$$
\eqalign{
\Bigl[\,D_{j_1,m_1;j_2,m_2}^{j_3,m_3}\,\Bigr]^2\,=&\,
\Bigl[\,C_{j_1,m_1;j_2,m_2}^{j_3,m_3}\,\Bigr]^4\,
\lambda(1)\,\,
{\cal P}^2(2j_1+2j_2+2j_3+1)\,
\times\cr\cr
&\times
\,\,\prod_{i=1}^{3}\,\,
{\lambda(4j_i+1)\,
{\cal P}^{2}(2j_1+2j_2+2j_3-4j_i)\over 
{\cal P}^2(4j_i+1)}\,\,. \cr}
\eqn\cien
$$
The result \cien\ constitutes the culmination of our
efforts. Its consequences and implications will be analyzed
in the next section. Before proceeding with this analysis
several remarks are in order. First of all, let us notice that
the structure constants $D_{j_1,m_1;j_2,m_2}^{j_3,m_3}$
depend on the $m_i$'s only through the corresponding
Clebsch-Gordan coefficients. This fact, which was to be
expected on general grounds, can be checked by studying some
other correlators, different from the one displayed in eq.
\cuno. In this calculation the representations \vcuatro\ and
\vcinco\ of the components of the multiplet of conjugate
primary fields must be used. We have verified in some of
these cases that, indeed, the result for the structure
constants is the one appearing in eq. \cien. Moreover, it is
interesting to point out that the $m_i$-independent factor in
the right-hand side of eq. \cien\ is symmetric under the
permutation of $j_1$, $j_2$ and $j_3$. Finally, let us notice
the close similarity of the result in  eq. \cien\ and the
value of the structure constants for the $sl(2)$ current
algebra (see refs. [\ZF, \Dotsenko]). In the next section this
similarity between the $sl(2)$ and ${\rm osp}(1\vert 2)$ 
cases will become more manifest.

\chapter{Fusion rules and connection with 
the superconformal minimal models}

With eq. \cien\ at our disposal we can obtain the selection
rules of the operator algebra, \ie\ the fusion rules, for the 
${\rm osp}(1\vert 2)$ current algebras. In order to determine
these rules we must characterize the values of $j_1$, $j_2$
and $j_3$ for which $D_{j_1,m_1;j_2,m_2}^{j_3,m_3}$ is 
non-vanishing. Let us first of all recall (see section 2)
that, when $k$ is a positive integer, 
only those representations of the affine ${\rm osp}(1\vert 2)$
superalgebra with isospins 
$$
j\,\le\,{k\over 2}\,,
\eqn\ctuno
$$
are allowed. Therefore, we shall assume that the constraint
\ctuno\ is satisfied by $j_1$, $j_2$ and $j_3$. It is
important to point out that when this happens a possible
divergence in \cien\ due to the terms 
$1\,/\,{\cal P}^2(4j_i+1)$ never takes place. We can
therefore concentrate ourselves on the zeros coming from the
other terms in \cien. Generally these zeros are generated
when a $\Gamma$-function evaluated at zero or a negative
integer appears in the denominator of eq. \cien. So, for
example, a detailed analysis of the factor 
${\cal P}^2(2j_1+2j_2+2j_3+1)$ using the definition \nsiete\
shows that it is non-vanishing only when $j_3$ satisfies the
condition:
$$
j_3\,\le\,k\,+\,{1\over 2}\,-\,j_1\,-\,j_2\,.
\eqn\ctdos
$$
Moreover, the term ${\cal P}^2(2j_1+2j_2-2j_3)$ is different
from zero  if 
$$
j_3\,\le\,j_1+j_2\,,
\eqn\cttres
$$
whereas, when \ctdos\ and \cttres\ are satisfied, the
requirement that the remaining factors 
${\cal P}^2(2j_2+2j_3-2j_1)$ and ${\cal P}^2(2j_1+2j_3-2j_2)$ 
 never vanish yields the condition:
$$
j_3\,\ge\,|j_1\,-\,j_2\,|\,.
\eqn\ctcuatro
$$
Notice that the inequalities \cttres\ and \ctcuatro\ do not
depend on the level $k$ and, in fact, are the same that
appear in the non-affine ${\rm osp}(1\vert 2)$ representation
theory. As  was mentioned before, they are automatically
incorporated in our approach when $j_3$ is given by eq.
\ciocho. On the other hand, eq. \ctdos\ is a non-trivial
constraint on the maximum value of $j_3$ which does depend on
the level $k$. It is elementary to verify the compatibility of
\ctdos\ with our general condition \ctuno. Indeed, by adding
\ctdos\ and \cttres\ one gets 
$j_3\,\le\,{k\over 2}\,+\,{1\over 4}$ which, as $k$ is
integer and $j_3$ can only take integer or half-integer
values, reduces to $j_3\,\le\,{k\over 2}$. Gathering all the
conditions we have found, we can write the 
${\rm osp}(1\vert 2)$ fusion rules as:

$$
[j_1]\,\times\,[j_2]\,=\,
\sum_{{j_3=|j_1-j_2|\atop}\atop 2(j_3-j_1-j_2)\,\in\,\ZZ}
^{{\rm min}\,(\,j_1+j_2\,,\,k+{1\over 2}-j_1-j_2\,)}
\,\,\,[j_3]\,\,.
\eqn\ctcinco
$$
Interestingly, the fusion rules \ctcinco\ coincide, 
for this $0<k\in \ZZ$ case, with the
ones found in ref. [\yudos] from the modular properties of the 
${\rm osp}(1\vert 2)$ characters. This coincidence is an
argument in favor of the validity in this case of the Verlinde
formula which, with a suitable reinterpretation, 
was used in [\yudos]
to arrive at eq. \ctcinco.

Let us now compare  the fusion rules \ctcinco\ with the ones
corresponding to the $N=1$ superconformal minimal models in
the Neveu-Schwarz (NS) sector. The $(p',p)$ superconformal
model has central charge:

$$
c_{p',p}\,=\,{3\over 2}\,[1-{2(p'-p)^2\over p'p}]\,\,.
\eqn\ctseis
$$
The primary
operators in the NS sector are labelled by two integers $m'$ and
$m$ which must be such that $1\le m'\le p'-1$, 
$1\le m\le p-1$ and $m'-m\in 2\ZZ$. The fusion rules for
these operators can be written as [\SCFT]:
$$
[\,m_1'\,,\,m_1\,]\,\times\,[\, m_2'\,,\,m_2\,]\,=
\!\!\!
\sum_{{m_3'=|m_1'-m_2'|+1\atop}\atop m_3'\in\,2\ZZ+1} 
^{{\rm min}\,(\,m_1'+m_2'-1\,,\,2p'-m_1'-m_2'-1\,)}
\,\,\,\,
\sum_{{m_3=|m_1-m_2|+1\atop}\atop m_3\in\,2\ZZ+1} 
^{{\rm min}\,(\,m_1+m_2-1\,,\,2p-m_1-m_2-1\,)}
\!\!\!\!\!\![\,m_3'\,,\,m_3\,]\,.
\eqn\ctsiete
$$
By inspecting \ctsiete, one concludes that these fusion rules
are equivalent to those of two independent 
${\rm osp}(1\vert 2)$ algebras, when the isospins $(j',j)$ and
levels $(k',k)$ of the latter are properly identified with the
quantum numbers $(m',m)$ and the $(p',p)$ parameters of the
$N=1$ superconformal field theory. This identification is:

$$
\eqalign{
&m'=4j'+1
\,\,\,\,\,\,\,\,\,\,\,\,\,\,
m=4j+1 \cr
&p'\,=2k'+3
\,\,\,\,\,\,\,\,\,\,\,\,\,\,
p=2k+3\,.\cr}
\eqn\ctocho
$$
One can check, using \ctocho, that the ranges of allowed
values  of $m$ and $m'$ 
(\ie\ $1\le m'\le p'-1$, $1\le m\le p-1$ and $m'-m\in 2\ZZ$) 
 correspond precisely to the ranges 
$j'\,\le \,k'/2$ and $j\,\le \,k/2$ for 
the ${\rm osp}(1\vert 2)$ isospins.

The relation between the ${\rm osp}(1\vert 2)$ current
algebras and  the $N=1$ superconformal theories goes beyond
the similarity of their fusion rules. Actually, the structure
constants for the products of thermal operators $\phi_{1,m}$
of the $N=1$ SCFT and those of the ${\rm osp}(1\vert 2)$
current algebras are closely related.  Let us denote by 
${\cal D}_{m_1,m_2}^{m_3}$ the structure constants appearing
in the correlator 
$<\phi_{1,m_1}(z_1)\phi_{1,m_2}(z_2)\phi_{1,m_3}(z_3)>$
and, for the $(p',p)$ N=1 minimal model, let $\tilde \rho$ be
defined as:
$$
\tilde \rho\,=\,{p'\over 2p}\,.
\eqn\ctnueve
$$
In order to express the constants ${\cal D}_{m_1,m_2}^{m_3}$
in a compact fashion, let us introduce the quantities
$$
\eqalign{
s_m\,\equiv&\,{1+\sum_{i=1}^3 \,\,m_i\over 4}\cr
Q(N)\,\equiv&\,\prod_{i=1}^N\,\,
{1\over (\,{1\over 2}\,-\,(2i+1)\tilde \rho\,)^2}\,,\cr}
\eqn\ctdiez
$$
in terms of which the function $\Lambda (m_1,m_2,m_3)$ is
defined as:
$$
\eqalign{
\Lambda (m_1,m_2,m_3)\,=&\,
2^{4\,(s_m\,-\,[s_m])}\,(-1)^{2s_m}\,\,
{Q^2\,(\,[\,s_m-1]\,)\over [s_m-{1\over 2}]!}\,\,\times\cr\cr
&\times\,\prod_{i=1}^3\,\,
{({m_i-1\over 2})!\over [{m_1+m_2+m_3-2m_i-1\over 4}]!\,\,
Q({m_i-1\over 4})}
\,\,.\cr}
\eqn\ctonce
$$
The value of the ${\cal D}_{m_1,m_2}^{m_3}$ constants has
been obtained in refs. [\Kita, \Zaugg]. With our notation the
result of these references can be written as:
$$
\eqalign{
\Bigl[\,{\cal D}_{m_1,m_2}^{m_3}\,\Bigr]^2\,=&\,
\Lambda(m_1,m_2,m_3)\,\,
\tilde\lambda(1)\,\,
\tilde{\cal P}^{\,2}(\,{m_1+m_2+m_3-1\over 2}\,)\,
\times\cr\cr
&\times
\,\,\prod_{i=1}^{3}\,\,
{\tilde\lambda(m_i)\,
\over \tilde{\cal P}^{\,2}(m_i)}\,\,
\tilde{\cal P}^{\,2}(\,{m_1+m_2+m_3-2m_i-1\over 2}\,)\,,
\cr}
\eqn\ctdoce
$$
where $\tilde\lambda(j)$ and $\tilde{\cal P}(j)$ are given by
eqs. \nseis\ and \nsiete\ with $\rho$ substituted by 
$\tilde\rho$. Notice that $\Lambda(m_1,m_2,m_3)$ never
vanishes and it is totally symmetric in $m_1$, $m_2$ and
$m_3$.  Moreover, comparing the
$\Gamma$-function terms in
\cien\ and \ctdoce\ one easily concludes that they are the
same if one identifies $\rho$ and $\tilde\rho$ and, as  was
done in eq. \ctocho, if we take $m_i=4j_i+1$. This
identification $\rho=\tilde\rho$  corresponds to taking
$2k+3\,=\,p/p'$, which is precisely the fractional level
required to construct the $N=1$ $(p',p)$ minimal
supersymmetric models from the hamiltonian reduction of the
${\rm osp}(1\vert 2)$ current algebra. It is interesting at
this point to recall that, between the structure constants of
the $sl(2)$ CFT and those of the thermal operators of the
minimal non-supersymmetric models, there exists a relation
similar to the one we have found here between the two sets of
structure constants of eqs. \cien\ and \ctdoce.

As  was pointed out above, the identification
$\rho\equiv\tilde\rho$ requires to consider a value of $k$
which is, in general, rational. For these rational values of
$k$ a larger class of ${\rm osp}(1\vert 2)$ admissible
representations must be considered (see sect. 2). Indeed, in
this case, the isospin $j$ can also take rational values.
Comparing with a similar situation in the $sl(2)$ algebra, it
is plausible to think that, in order to deal with this more
general case, one must consider a formalism in which new
variables are introduced to represent the action of the
superalgebra. For the $sl(2)$ Lie algebra the primary fields
in this new formalism depend on  the space-time coordinate
$z$ and on a new bosonic coordinate $x$. In the 
${\rm osp}(1\vert 2)$ case it is natural to introduce, in
addition, a new Grassmann variable $\theta$ in such a way
that the generalized primary fields are:
$$
\varphi(z,x,\theta)\,=\,(\,1\,+\,x\chi\,
+\,\theta\psi)^{2j}\,e^{-2ij\alpha_0\,\phi}\,\,.
\eqn\cttrece
$$
Notice that, for integer or half-integer $j$,
$\varphi(z,x,\theta)$ can be expanded in a finite sum
involving different powers of $x$ and $\theta$. The
coefficients in this power sum are precisely the fields
$\Phi_j^m$ defined in \diez. In fact, the terms that do not
contain the Grassmann variable $\theta$ correspond to
$\Phi_j^m$ with $j-m\in\ZZ$, whereas those that contain
$\theta$, and therefore also the field $\psi$, can be
identified with the $\Phi_j^m$ operators with 
$j-m\in\ZZ+{1\over 2}$. For general $j$, the OPE's of the  
${\rm osp}(1\vert 2)$ currents with the field 
$\varphi(z,x,\theta)$ can be written as:
$$
\eqalign{
H(z)\,\varphi(w,x,\theta)\,=&\,
{D_3^j\,\varphi(w,x,\theta)\over z-w}\cr
J_{\pm}(z)\,\varphi(w,x,\theta)\,=&\,
{D_{\pm}^j\,\varphi(w,x,\theta)\over z-w}\cr
j_{\pm}(z)\,\varphi(w,x,\theta)\,=&\,
{d_{\pm}^j\,\varphi(w,x,\theta)\over z-w}\,\,,\cr}
\eqn\ctcatorce
$$
where $D_3^j$, $D_{\pm}^j$ and $d_{\pm}^j$ are the following
differential operators on the variables $x$ and $\theta$:
$$
\eqalign{
D_3^j\,=&\,-x\partial_x\,-\,{1\over 2}\,\theta\,
\partial_{\theta}\,+\,j\cr
D_+^j\,=&\,-x^2\partial_x\,+\,2jx\,-\,\theta x
\partial_{\theta}\cr
D_-^j\,=&\,\partial_x\cr
d_+^j\,=&\,x\partial_{\theta}\,+\,\theta x\partial_x\,
-\,2j\theta\cr
d_-^j\,=&\,\partial_{\theta}\,+\,\theta\partial_x\,\,.\cr}
\eqn\ctquince
$$
It is straightforward to verify that the operators \ctquince\
can be used to represent the ${\rm osp}(1\vert 2)$ algebra
(this is the so-called isotopic representation).

Let us finally point out that, when the isospin $j$ is
not integer or half-integer, the charge $Q$ given in eqs.
\vsiete\ and \vocho\ is unable to screen a general correlator.
We must thus find, as it happens in the $sl(2)$ CFT, a
second screening operator. This is not difficult and, in
fact, one can prove that the OPE's of the 
${\rm osp}(1\vert 2)$ currents with the field
$$
\widehat S\,=\,w^{-(k+2)}\,\,(\,\bar \psi\,-\,w\psi\,)\,
e^{\,-{i\over \alpha_0}\,\phi}\,,
\eqn\ctseis
$$
have only total derivatives. Therefore 
$\oint\,dz\widehat S(z)$ can be taken as a screening charge.
Notice that $\widehat S(z)$ is non-local (this also happens
for $sl(2)$) and, thus, one must give a prescription to compute
correlators involving it. Presumably one could apply, for this
purpose, the fractional calculus technique of ref.
\REF\peter{J. L. Petersen, J. Rasmussen and M. Yu
\journal\np&B457(95)309\journal\np&B457(95)343.} [\peter].

\chapter{Concluding remarks}

Let us recapitulate our main results. We have studied 
the free field realization of the ${\rm osp}(1\vert 2)$
current algebra. We have given a representation of the
primary fields of the theory which allows to compute the
conformal blocks of the model. We have been able to define a
set of selection rules for the computation of the Fock space
expectation values, which are such that incorporate the basic
features of the ${\rm osp}(1\vert 2)$ representation theory.
We have performed in detail the analysis of the four
point-functions. Our goal in this study was to obtain 
the structure constants appearing in  the operator algebra of
the theory. 

The technique we have used has been successfully
employed previously  for the minimal models [\DF, \Kita,
\Zaugg] and for the
$sl(2)$ current algebras[\Dotsenko]. The computation of the
structure constants requires the evaluation of the
normalization integrals of the blocks. This is, in fact, the
greatest technical difficulty of this approach. In our case,
the integrals needed  can be computed in some
cases by reducing them  to known
results, and by imposing some consistency conditions in  other
situations. Although the expression of these integrals might
appear cumbersome, the final result
\cien\ for the structure constants is rather simple and,
actually, very similar to the $sl(2)$ case. This is so because
some remarkable simplifications occur in the intermediate
steps of the calculation.

Once the structure constants are known one can perform an
analysis to determine when they vanish. The outcome of this
study are the fusion rules \ctcinco\ of the model. The
comparison of these rules with the ones corresponding to the
minimal superconformal theories can help to shed light on the
relation between them. For example, the
interpretation of the fusion rules for the $N=1$ minimal
superconformal models as given by two independent 
${\rm osp}(1\vert 2)$ algebras is reminiscent of a similar
relationship between the minimal Virasoro models and the
$sl(2)$ current algebras. This latter relation can be easily
understood when one constructs the minimal Virasoro models as
$sl(2)$ coset theories. Our result suggests the existence of a
similar construction relating the ${\rm osp}(1\vert 2)$
superalgebras and the $N=1$ minimal superconformal models.

Obviously, many aspects of our free field construction remain
to be explored. For example, one should be able to obtain the
full duality structure of the ${\rm osp}(1\vert 2)$ theory by
applying the ideas of ref. 
\REF\gomez{C. Gomez and G.
Sierra\journal\pl&B240(90)149\journal\np&B352(91)791.}
 [\gomez] to our approach. One expects
that a quantum deformation of ${\rm osp}(1\vert 2)$ 
\REF\saleur{P. Kulish and N.
Reshetikhin\journal\lmp&18(89)143;
H. Saleur\journal\np&B336(90)363.}
 [\saleur] will
show up as the result of this analysis. It is also likely that 
the fusion rules \ctcinco\ could be obtained directly by using
the cohomological methods of ref. 
\REF\felder{G. Felder\journal\np&B317(89)215; D. Bernard
and G. Felder\journal\cmp&127(90)145.}[\felder].

One should be 
able to incorporate to our formalism the fractional
isospins that appear when the level $k$ is not integer. The
understanding of the conformal blocks for these
representations is essential if one wants to implement a
hamiltonian reduction procedure,  relating the 
${\rm osp}(1\vert 2)$ theory with the supersymmetric theories,
and, eventually, with the two-dimensional supergravity and the
non-critical superstrings. In this respect, the isotopic
approach introduced at the end of section 5 should be
relevant. Within this approach, it is natural to assemble the
two internal coordinates $x$ and $\theta$ into an isotopic
superspace coordinate $X=(x,\theta)$. On the other hand,  
\REF\Furlan{P. Furlan et al. \journal\pl&B267(91)63
\journal\np&B394(93)665.}
 it was argued in ref. [\Furlan] that the hamiltonian reduction
of the $sl(2)$ theory can be performed by identifying the
isotopic and the space-time coordinates, \ie\ by putting
$x=z$. Trying to generalize this result to the ${\rm
osp}(1\vert 2)$ superalgebra,  one is led to think that one
should supplement the ${\rm osp}(1\vert 2)$ currents with some
supersymmetric partners. The resulting model would be a
Kac-Todorov 
\REF\KT{V.G. Kac and I.T. Todorov\journal\cmp&102(85)337;
P. Di Vecchia, V.G. Knizhnik, J.L. Petersen and P.
Rossi \journal\np&B253(85)701.}
 [\KT] system
for the ${\rm osp}(1\vert 2)$ Lie superalgebra, for which a
natural superspace coordinate $Z=(z,\eta)$ can be defined.
It would be very interesting to investigate if the
identification of $Z$ and $X$ could be used to generate the
conformal blocks for the $N=1$ superconformal models from
those of the ${\rm osp}(1\vert 2)$ system. The addition of
extra fields to the  ${\rm osp}(1\vert 2)$ current algebra has
been considered in other approaches to hamiltonian reduction
\REF\kura{T. Kuramoto\journal\np&B411(94)821.}
[\kura].

Let us finally point out that the methods employed here can
be applied, in principle, to other Lie superalgebras. The
simplest of these superalgebras is ${\rm osp}(2\vert 2)$,
which is related to the $N=2$ superconformal symmetry
[\bershadsky]. The ${\rm osp}(2\vert 2)$ representation
theory is well established [\Pais] and the free field
representation of the corresponding current algebra has been
given in ref. [\bershadsky].

\ack
We are grateful to L. Alvarez-Gaume, J.M.F. Labastida 
and P.M. Llatas for discussions. We thank J. M. Isidro 
and J. Mas for a critical reading of the manuscript. 
This work was supported in part by DGICYT under
grant PB93-0344, and by CICYT under grant  AEN96-1673.

\Appendix A

The ${\rm osp}(1\vert 2)$ Lie superalgebra contains three
bosonic generators $T_{\pm}$ and  $T_3$, which form the
Lie algebra $sl(2)$, together with two fermionic
generators $F_{\pm}$. The (anti)commutators that define
the ${\rm osp}(1\vert 2)$ superalgebra are:
$$
\eqalign{
&[T_3\,,\,T_{\pm}]\,=\,\pm T_{\pm}
\,\,\,\,\,\,\,\,\,\,\,\,\,\,\,\,\,\,\,\,\,
\,\,\,\,\,\,\,\,\,\,\,\,\,\,\,\,\,\,\,\,\,
[T_{+}\,,\,T_{-}]\,=\,2T_3\cr
&[T_3\,,\,F_{\pm}]\,=\,\pm {1\over 2} F_{\pm}
\,\,\,\,\,\,\,\,\,\,\,\,\,\,\,\,\,\,\,\,\,
\,\,\,\,\,\,\,\,\,\,\,\,\,\,\,\,
\{F_{\pm}\,,\,F_{\pm}\}\,=\,\pm 2 T_{\pm}\cr
&\{F_{+}\,,\,F_{-}\}\,=\, 2 T_{3}
\,\,\,\,\,\,\,\,\,\,\,\,\,\,\,\,\,\,\,\,\,
\,\,\,\,\,\,\,\,\,\,\,\,\,\,\,\,\,\,\,\,\,
[T_{\pm}\,,\,F_{\pm}]\,=\,0\cr
&[T_{\pm}\,,\,F_{\mp}]\,=\,-F_{\pm}\,\,.\cr}
\eqn\apauno
$$
Notice that the bosonic generators $T_{\pm}$ and $T_3$
correspond to the currents $J_{\pm}(z)$ and $H(z)$
respectively, whereas the fermionic operators $F_{\pm}$
correspond to the currents $j_{\pm}(z)$. It is not
difficult to prove, using the relations \apauno, that the
operator
$$
C_2\,=\,T_3^2\,+\,{1\over 2}\,[\,T_-T_+\,+\,T_+T_-\,]\,+\,
{1\over 4}\,[\,F_-F_+\,-\,F_+F_-\,]\,,
\eqn\apados
$$
commutes with all the ${\rm osp}(1\vert 2)$ generators.
Actually, $C_2$ is the quadratic Casimir operator of the 
${\rm osp}(1\vert 2)$ superalgebra. Using the
(anti)commutators \apauno\ we can reexpress $C_2$ as:
$$
C_2\,=\,T_3^2\,+\,{1\over 2}\,T_3\,+\,T_-T_+\,+\,
{1\over 2}\,F_-F_+\,.
\eqn\apatres
$$

Let us now study [\Pais] the matrix 
representations of the algebra
\apauno. The representation theory of 
${\rm osp}(1\vert 2)$ has many features that are similar to
those of  the $sl(2)$ Lie  algebra. As in this
latter case, we shall represent the Cartan generator $T_3$
by a diagonal operator and, thus, we can label  the states
of  our vector space basis by their $T_3$ eigenvalues,
which are the weights of the representation. It is a simple
exercise to check from \apauno\ that the $T_{+}$ ($T_{-}$)
operator raises (lowers) the eigenvalue of the $T_3$
eigenstates in one unit without changing its Grassmann
parity, whereas  $F_{+}$ ($F_{-}$) increases (decreases)
the $T_3$ eigenvalue of these states in one-half unit and
changes their statistics. The finite dimensional
irreducible representations ${\cal R}_j$ of 
${\rm osp}(1\vert 2)$ are characterized by the value $j$
of their highest weight, which can be integer or
half-integer. In close parallel with the $sl(2)$ case, we
shall call $j$  the isospin of the representation.
If we denote the highest weight vector by $|j,j>$, it is
evident that it must satisfy: 
$$
T_+\,|j,j>\,=\,F_+\,|j,j>\,=\,0\,.
\eqn\apacuatro
$$
A general basis state for the representation ${\cal R}_j$
will be denoted by $|j,m>$, $m$ being the $T_3$
eigenvalue. The quadratic Casimir operator $C_2$ acts on
these states as a multiple of the unit operator. It is 
easy to obtain its value by computing $C_2|j,j>$ using
\apacuatro. One immediately gets:
$$
C_2\,|j,m>\,=\,j\,(\,j+{1\over 2}\,)\,|j,m>\,.
\eqn\apacinco
$$

Using eqs. \apacinco, \apados\ and the defining relations
of the algebra (eq. \apauno), it is not difficult [\Pais] to
find the matrix elements of the different generators in
the representation  ${\cal R}_j$. The result that one
finds for the bosonic operators is:
$$
\eqalign{
&T_{3}\,|j,m>\,=\,m|j,m>\cr\cr
&T_{\pm}\,|j,m>\,=\,\sqrt{\,[j\mp m]\,[j\pm m+1]}\,
|j,m\pm 1>\,\,,\cr}
\eqn\apaseis
$$
where, as in the main text, $[x]$ represents the integer
part of a non-negative integer or half-integer $x$. The
matrix elements of the odd operators $F_{\pm}$ are:
$$
F_{\pm}\,|j,m>\,=\,\cases{
-\sqrt{j\mp m}\,|j,m\pm {1\over 2}>&if $j-m\in \ZZ$\cr\cr
\mp\sqrt{j\pm m+{1\over 2}}\,|j,m\pm {1\over 2}>
&if $j-m\in\ZZ+{1\over 2}$.}
\eqn\apasiete
$$

From \apaseis\ and \apasiete\ one easily concludes that,
indeed, when $j$ is integer or half-integer ${\cal R}_j$
is finite dimensional. In fact, in this case, only the
states $|j,m>$ with 
$m=-j, -j+{1\over 2},\,\cdots\,,j-{1\over 2}, j$ are
connected by the action of the ${\rm osp}(1\vert 2)$
generators. The dimension of  ${\cal R}_j$ is thus:
$$
{\rm dim}\,(\,{\cal R}_j\,)\,=\,4j+1\,.
\eqn\apaocho
$$
Moreover, eq. \apaseis\ shows that for $j\not=0$  
${\cal R}_j$ decomposes under the even part  of the
superalgebra into two $sl(2)$ multiplets with isospins
$j$ and $j-{1\over 2}$. The members of these multiplets
are the states $|j,m>$ with $j-m\in\ZZ$ and 
$j-m\in\ZZ+{1\over 2}$ respectively. These two $sl(2)$
multiplets are connected by the action of $F_{\pm}$ and
thus they have opposite statistics. Actually, an 
 ${\rm osp}(1\vert 2)$ representation ${\cal R}_j$ is
completely characterized if, together with its isospin $j$,
we also give the statistics of its highest weight state 
$|j,j>$. The Grassmann parity of $|j,j>$ will be denoted
by $p(j)$ ($p(j)=0,1$). We will say that the
representation is even (odd) when $|j,j>$ is
bosonic(fermionic), \ie\ when $p(j)=0$ ($p(j)=1$).

For Lie superalgebras it is possible to define a
generalized adjoint operation [\Pais], denoted by $\ddagger$,
such that for any operator $A$ and any two states
$\alpha$ and $\beta$  one has:
$$
<A^\ddagger \alpha | \beta>\,=\,(-1)^{p(A)p(\alpha)}\,\,
<\alpha|A\beta>\,,
\eqn\apanueve
$$
where $p(A)$ and  $p(\alpha)$ denote respectively the
Grassmann parities of the operator $A$ and the state
$\alpha$. We will say that $A^{\ddagger}$ is the
superadjoint of $A$. From the property \apanueve, one can
verify that:
$$
(AB)^{\ddagger}\,=\,(-1)^{p(A)p(B)}\,\,
B^{\ddagger}\,A^{\ddagger}\,\,.
\eqn\apadiez
$$
It is not difficult to obtain the explicit form of the
superadjoint operation for the  ${\rm osp}(1\vert 2)$
generators. In fact, by requiring compatibility of this
operation with the (anti)commutators \apauno, one can
easily establish that $T_{\pm}^{\ddagger}\,=\,T_{\mp}$ and 
$T_3^{\ddagger}\,=\,T_3$, as expected, while the rule for
the fermionic generators is:
$$
F_{+}^{\ddagger}\,=\,\eta\,F_{-}
\,\,\,\,\,\,\,\,\,\,\,\,\,\,\,\,\,\,\,\,\,
F_{-}^{\ddagger}\,=\,-\eta\,F_{+}\,\,,
\eqn\apaonce
$$
where $\eta$ can take the values $\pm 1$. Notice that,
independently of $\eta$, 
$((F_{\pm})^{\ddagger})^{\ddagger}\,=\,-F_{\pm}$. The
actual value of $\eta$ can be determined, as a
consequence of eq. \apanueve, from the norm of the basis
states and the parity of the representation. Let us
suppose that 
$<j,m|j,m>\,=\,\epsilon (\epsilon\,')$ if $j-m$ is
integer (half-integer), where $\epsilon$ and
$\epsilon\,'$ can take the values $\pm 1$. Putting in
eq. \apanueve\ $\alpha\,=\,|j,j>$,  
$\beta\,=\,|j,j-{1\over 2}>$ and $A=F_{+}$, and taking eq.
\apasiete\ into account, one gets:
$$
\eta\,=\,(-1)^{p(j)}\,\epsilon\epsilon\,'\,\,.
\eqn\apadoce
$$
We shall conventionally choose $\eta=1$, which means that 
$F_{+}^{\ddagger}\,=\,F_{-}\,\,\,\,$ and    
$\,\,F_{-}^{\ddagger}\,=\,-F_{+}$. For even
representations this election implies that 
$\epsilon\epsilon\,'\,=\,+1$, whereas, on the contrary, 
$\epsilon\epsilon\,'$ must be negative for odd
representations (see eq. \apadoce). According to this
result we shall take $\epsilon\,=\,\epsilon\,'\,=\,+1$(
$\epsilon\,=-\,\epsilon\,'\,=\,+1$) for even(odd)
representations and, thus, only the states $|j,m>$ with 
$p(j)=1$ and $j-m\in \ZZ+{1\over 2}$ will have negative
norm.

Let us now consider  the tensor product of two
representations. By using the well-known methods of
angular momentum theory, one can easily convince oneself
that the coupling of isospins $j_1$ and $j_2$ gives rise
to isospins $j_3\,=\,|j_1-j_2|, |j_1-j_2|+{1\over 2}, 
\cdots, j_1+j_2$. Actually, one can write the tensor
product decomposition of 
${\cal R}_{j_1}\,\otimes {\cal R}_{j_2}$ as:
$$
{\cal R}_{j_1}\,\otimes {\cal R}_{j_2}\,=\,
\bigoplus_{{j_3=|j_1-j_2|\atop}\atop 2(j_3-j_1-j_2)\,\in\,\ZZ}
^{j_1+j_2}
\,{\cal R}_{j_3}\,\,.
\eqn\apatrece
$$
Furthermore, the parity of the representations 
${\cal R}_{j_3}$ in the right-hand of eq. \apatrece\ is:
$$
p(j_3)\,=\,p(j_1)\,+\,p(j_2)\,+\,2(j_1+j_2-j_3)
\,\,\,\,\,\,\,\,\,\,\,\,\,\,\,\,\,\,\,\,
{\rm mod}\,(2)\,\,.
\eqn\apacatorce
$$
As usual, the states $|\,j_3,m_3\,>$ can be obtained from
those of ${\cal R}_{j_1}\,\otimes {\cal R}_{j_2}$ by
means of the ${\rm osp}(1\vert 2)$ Clebsch-Gordan
coefficients $C_{j_1,m_1;j_2,m_2}^{j_3,m_3}$:
$$
|\,j_3,m_3\,>\,=\,\sum_{m_1,m_2}\,
C_{j_1,m_1;j_2,m_2}^{j_3,m_3}\,\,\,\,
|j_1,m_1>\otimes\,|j_2,m_2>\,\,.
\eqn\apaquince
$$

In ref. [\Pais],  the
$C_{j_1,m_1;j_2,m_2}^{j_3,m_3}$ have been computed in
terms of the $sl(2)$ Clebsch-Gordan coefficients. Let us
denote the latter by 
${\widehat C}_{j_1,m_1;j_2,m_2}^{\,\,j_3,m_3}$. In our
calculation of the structure constants of the 
${\rm osp}(1\vert 2)$ current algebra, we shall only need
the value of the $C_{j_1,m_1;j_2,m_2}^{j_3,m_3}$ for 
$j_1-m_1\in \ZZ$ and $j_2-m_2\in \ZZ$. In this case the
result given in ref. [\Pais] can be written as:
$$
C_{j_1,m_1;j_2,m_2}^{j_3,m_3}\,\,=\,\,
\cases{\sqrt{{j_1+j_2+j_3+1\over 2j_3+1}}\,\,\,
{\widehat C}_{j_1,m_1;j_2,m_2}^{\,\,j_3,m_3}
&if $j_3-m_3\in \ZZ$\cr\cr
(-1)^{p(j_1)+1}\,
\sqrt{{j_1+j_2-j_3+{1\over 2}\over 2j_3}}\,\,\,
{\widehat C}_{j_1,m_1;j_2,m_2}^{\,\,j_3-{1\over 2},m_3}
&if $j_3-m_3\in \ZZ+{1\over 2}\,.$}
\eqn\apadseis
$$
Let us particularize eq. \apadseis\ to the situation in
which $m_1=j_1$ and $m_2=-j_2$. The $sl(2)$
Clebsch-Gordan coefficients are:
$$
{\widehat C}_{j_1,j_1;j_2,-j_2}^{\,\,j,j_1-j_2}\,\,=\,\,
\sqrt{2j+1}\,\,
\Bigl[\,
{(2j_1)!\,(2j_2)!\over (j_1+j_2+j+1)!\,(j_1+j_2-j)!}
\Bigr]^{1/2}\,\,.
\eqn\apadsiete
$$
Substituting  this result for $j=j_3$
and $j=j_3-{1\over 2}$  in eq. \apadseis, 
the following expression for 
$C_{j_1,j_1;j_2,-j_2}^{j_3,j_1-j_2}$ is obtained:
$$
C_{j_1,j_1;j_2,-j_2}^{j_3,j_1-j_2}\,=\,
(-1)^{2(p(j_1)+1)(j_1+j_2-j_3)}\,\,
\Bigl[\,
{\,(2j_1)!\,(2j_2)!\over
([j_1+j_2+j_3+{1\over 2}])!\,
([j_1+j_2-j_3])!}\,
\Bigr]^{1/2}\,\,,
\eqn\apadocho
$$
which is precisely the result needed in section 4 (see
eq. \nnueve).

\Appendix B

In this appendix we will evaluate the multiple 
integrals needed to
obtain the expression of the ${\rm osp}(1|2)$ 
structure constants.
In general, the integrals to compute will involve 
functions of $N$
variables $t_1,t_2, \cdots, t_N$. The domain of 
integration will
be the subset of the $N$-dimensional unit 
hypercube defined by 
$1\ge t_1\ge \cdots\ge t_{N-1}
\ge t_N\ge 0$. If $f(\,\{t_i\}\,)$
is the function to integrate, we shall adopt the following
notation for these ordered integrations:
$$
\oint\,\,f(\,\{t_i\}\,)\,\equiv\,
\int_0^1\,dt_1\,\cdots\,\int_0^{t_{N-1}}\,dt_N\,
\,\,f(\,\{t_i\}\,).
\eqn\apbuno
$$
First of all, let us consider the class of integrals 
defined by: 
$$
\eqalign{
I_n^m\,(a,b,\rho)\,\equiv\,\oint
&[\,(t_1\cdots t_m)^{a+1}\,(t_{m+1}\cdots t_n)^{a}\,+\,
{\rm permutations}\,]\,\times\cr
&\times \prod_{i=1}^n\,(\,1-t_i\,)^b\,
\prod_{i<j}^n\,(\,t_i\,-\,t_j\,)^{2\rho}\,,\cr}
\eqn\apbdos
$$
where $0\le m\le n$. In eq. \apbdos\ the dimensionality of 
the integration domain
is $N=n$ and, in the integrand, $m$ of the integration
variables are raised to the power 
$a+1$ whereas the exponent
of $n-m$ of them is $a$. A sum over all the possible
elections of these $n$ and $n-m$  variables is performed in
order to make the integrand in \apbdos\ completely
symmetric in its arguments. The result 
of the integrals $I_n^0\,(a,b,\rho)$
and $I_n^n\,(a,b,\rho)$  has been given by Dotsenko and
Fateev [\DF] as a product of Euler $\Gamma$-functions. It is
 easy to get an expression interpolating between these
two extreme values of $m$. Actually we are going to argue
that the integrals $I_n^m\,(a,b,\rho)$ are given by:
$$
\eqalign{
I_n^m\,(a,b,\rho)\,=&\,{n \choose m}\,\lambda_n(\rho)\,
\prod_{i=0}^{n-m-1}\,
{\Gamma (1+a+i\rho)\,\Gamma (1+b+i\rho)\over
\Gamma (2+a+b+(n-1+i)\rho)}\,\times\cr
&\times \prod_{i=n-m}^{n-1}\,
{\Gamma (2+a+i\rho)\,\Gamma (1+b+i\rho)\over
\Gamma (3+a+b+(n-1+i)\rho)}\,,\cr}
\eqn\apbtres
$$
where the function $\lambda_n(\rho)$ is:
$$
\lambda_n(\rho)\,=\,\prod_{i=1}^n\,
{\Gamma (i\rho)\over\Gamma (\rho)}\,.
\eqn\apbcuatro
$$
Notice that, in the last $m$ factors of the right-hand
side of \apbtres,  $a$ is shifted in one unit with respect
to the first $n-m$ ones. As a first check of eq. \apbtres, 
let us consider the case $\rho=0$. When $\rho$ vanishes
the multiple integral in eq. \apbdos\ decouples into $n$
one-dimensional integrals whose expression is given in
terms of the Euler beta function. It is straightforward to
verify that the value of $I_n^m\,(a,b,0)$ so obtained
coincides with the one dictated by \apbtres\ when
$\rho=0$. As a more restrictive test of \apbtres,  we are
going to get a functional relation that the integrals
$I_n^m\,(a,b,\rho)$ must satisfy. This  
relation can be obtained directly from the definition
\apbdos. Suppose that we use in this equation the identity 
$t_i^{a+1}\,=\,t_i^{a}\,-\,t_i^{a}(\,1-t_i\,)$ in all the
variables whose exponent is $a+1$. After changing
variables as $t_i\rightarrow 1-t_{n+1-i}\,$ for
$i=1,\cdots,n$, we arrive at the following relation:
$$
I_n^{\,m}\,(a,b,\rho)\,=\,\sum_{p=0}^m\,(-1)^{m-p}\,
{p+n-m\choose p}\,I_n^{\,m-p}\,(b,a,\rho)\,.
\eqn\apbcinco
$$
For $m=0$ eq. \apbcinco\ simply states that
$I_n^0\,(a,b,\rho)$ is symmetric under the interchange of
$a$ and $b$. Notice that this property is crucial in the
Dotsenko-Fateev derivation[\DF]. For $m>0$ eq. 
\apbcinco\ relates the integrals
\apbdos\ to functions of the same kind with
lower values of their upper index $m$ and with $a$ and $b$
exchanged. Using the elementary properties of the
$\Gamma$-function, it is not difficult to verify that our
result
\apbtres\ satisfies eq. \apbcinco. In order to get a
general proof of this statement, the binomial identity
\ocuatro\ is very useful. 

As a final check of eq. \apbtres, let us compute the
integrals \apbdos\ for $m=n-1$ by means of the following
trick. First of all, we define the functions 
$K_n(a,b,\rho,x)$ depending on an additional variable $x$:
$$
K_n(a,b,\rho,x)\,\equiv\,\int_0^{x}\,
dt_1\cdots\int_0^{t_{n-1}}\,dt_{n}\,
\prod_{i=1}^n\,t_i^a\,(x-t_i)^{b+1}\,
\prod_{i<j}^n\,(t_i-t_j)^{2\rho}\,.
\eqn\apbseis
$$
It is evident from their definition that the integrals 
$K_n(a,b,\rho,x)$ reduce to the functions written in eq.
\apbdos\ when the variable $x$ is equal to one. Namely,
one has:
$$
K_n(a,b,\rho,1)\,=\,I_n^{\,0}(a, b+1, \rho)\,.
\eqn\apbsiete
$$
Moreover, the dependence of $K_n(a,b,\rho,x)$ on $x$ can
be easily extracted by performing a rescaling
$t_i\rightarrow xt_i$ of the integration variables in eq.
\apbseis. One easily arrives at:
$$
K_n(a,b,\rho,x)\,=\,x^{2n+na+nb+n(n-1)\rho}\,\,
I_n^{\,0}(a, b+1, \rho).
\eqn\apbocho
$$
Let us now calculate the derivative of $K_n(a,b,\rho,x)$
with respect to $x$ at the point $x=1$. Computing it 
 directly from the definition \apbseis, one
gets:
$$
{\partial K_n(a,b,\rho,x)\over \partial x}|_{x=1}\,=\,
(b+1)\,I_n^{\,n-1}\,(b,a,\rho)\,.
\eqn\apbnueve
$$
The left-hand side of eq. \apbnueve\ can also be obtained
from the $x$ dependence displayed in eq. \apbsiete.
Comparing both ways of computing this derivative, one
arrives at the result:
$$
I_n^{\,n-1}\,(a,b,\rho)\,=\,n\,\,{2+a+b+(n-1)\rho\over
1+a}\,\, I_n^{\,0}(b, a+1, \rho)\,,
\eqn\apbdiez
$$
which gives the integrals \apbdos\ for $m=n-1$ in terms
of those with $m=0$. Using the Dotsenko-Fateev result for 
$I_n^{\,0}(a, b, \rho)$, one easily proves that the values
of $I_{\,n}^{n-1}\,(a,b,\rho)$ given in eqs. \apbdiez\ and
\apbtres\ coincide.

We are now going to study a family of integrals that
appear directly in our evaluation of the ${\rm osp}(1|2)$
operator algebra. They are the $2n$-dimensional integrals 
$J_{2n}^{m}\,(a\,,b\,,\rho\,)$, with $0\le m\le n$, defined
as:
$$
\eqalign{
J_{2n}^{\,m}\,(a\,,b\,,\rho\,)\,\equiv&\,\oint\,
\{\,(t_1\cdots t_n)^a\,
[\,(t_{n+1}\cdots t_{n+m})^{a+1}\,
(t_{n+m+1}\cdots t_{2n})^{a}\,
+\,{\rm  permutations}\,]\times\cr
&\times\,<\psi(t_1)\cdots \psi(t_n)\bar\psi(t_{n+1})
\cdots\bar\psi(t_{2n})>\,+\,{\rm  permutations}\,\}\times\cr
&\times\,\prod_{i=1}^{2n} (1-t_i)^b
\prod_{i<j}^{2n}\,(\,t_i\,-\,t_j\,)^{2\rho}\,.\cr}
\eqn\apbonce
$$
Notice the close similarity between the definitions  
 \apbdos\ and \apbonce. The main difference between
  $I_n^m$ and $J_{2n}^{m}$ is the presence in
the latter of a fermionic correlator involving the Dirac
fields $\psi$ and $\bar \psi$. Our conventions for the
normalization of these fields are the same as those used
in section 2 (see eq. \uno). In eq. \apbonce\ $m$ of the 
arguments of the $\bar\psi$ fields appear raised to the
power $a+1$. A double symmetrization of the integrand of
\apbonce\ is performed. First of all, one must sum over
all the possible elections of the $m$ variables among the $n$
arguments of the $\bar \psi$ fields inserted in the
correlator. Secondly, one must sum over all the possible
locations of the fields $ \psi$ and $\bar \psi$ inside
the vacuum expectation value. It is important to point
out that the arguments of the fields appearing in these
correlators are ordered as the integration limits, \ie\ 
when  $i<j$, 
the field with argument $t_i$ is always to the left of
those with arguments $t_j$. 

The value of the integrals $J_{2n}^m$ can be obtained
following the same steps that led us to \apbtres. The
expression that one arrives at is:
$$
\eqalign{
J_{2n}^{m}\,(a\,,b\,,\rho\,)\,=&\,{n\choose m}\mu_{2n}(\rho)\,
\prod_{i=0}^{2n-2m-1}\,
{\Gamma(1+a+i(\rho-{1\over 2})+[{i\over 2}])\,
\Gamma(1+b+i(\rho-{1\over 2})+[{i\over 2}])\over
\Gamma(1+a+b+n+(\rho-{1\over 2})(2n-1+i)+[{i\over 2}])}\times\cr\cr
&\times\prod_{i=2n-2m}^{2n-1}\,
{\Gamma(1+a+i(\rho-{1\over 2})+[{i+1\over 2}])\,
\Gamma(1+b+i(\rho-{1\over 2})+[{i\over 2}])\over
\Gamma(1+a+b+n+(\rho-{1\over 2})(2n-1+i)+[{i+1\over 2}])}
\,,\cr}
\eqn\apbdoce
$$
where the function $\mu_N(\rho)$ is given by:
$$
\mu_N(\rho)\,=\,\prod_{i=1}^N\,
{\Gamma (i(\rho+{1\over 2})\,-\,[{i\over 2}])
\over \Gamma (\rho+{1\over 2})}\,.
\eqn\apbtrece
$$
In eqs. \apbdoce\ and \apbtrece\ $[{i\over 2}]$
represents the integer part of ${i\over 2}$ for any
positive integer $i$. In order to verify the correctness
of eqs. \apbdoce\ and \apbtrece, let us consider some
particular cases. Let us first take $m=0$ in eq.
\apbdoce. For this value of $m$ the fermionic
correlators appearing in the definition of $J_{2n}^m$ are
all multiplied by the same factor. It turns out that the
combination of vacuum expectation values of products of
$\psi$ and $\bar\psi$ appearing in  $J_{2n}^0$ can be
put  as a single correlator of a new fermionic field.
Indeed, let $\lambda(t)$ be a fermionic Majorana field
normalized in such a way that its basic OPE is 
$\lambda(t_1)\lambda(t_2)\,=\,(\,t_1-t_2\,)^{-1}$. It can
be easily proved that:
$$
\eqalign{
<\psi(t_1)&\cdots \psi(t_n)\bar\psi(t_{n+1})
\cdots\bar\psi(t_{2n})>\,+\,{\rm  permutations}\,=\cr\cr
&=\,2^n\,<\lambda(t_1)\cdots \lambda(t_n)\lambda(t_{n+1})
\cdots\lambda(t_{2n})>\,.
\cr}
\eqn\apbcatorce
$$

Integrals of the type studied above,\ie\ with a correlator of
Majorana fields in the integrand,  appear in the
Feigin-Fuchs calculation of the structure constants of the
minimal supersymmetric models[\Kita, \Zaugg]. In fact, a
general expression for these integrals has been given in ref.
[\Kita]. After taking eq. \apbcatorce\ into account, it can be
seen that our expression \apbdoce\ for $m=0$ is in
agreement with the result of ref. [\Kita].

It would be interesting to find a particular value of
$\rho$ for which the $2n$-dimensional integral \apbonce\
decouples. Notice that, due to the presence of the
fermionic correlator, this decoupling does not occur now
at $\rho=0$. In order to find the new decoupling point,
let us study for a while the fermionic correlator. Using
the two-point function for the $\psi$ and $\bar\psi$
fields (see eq. \uno) and taking the anticommutative
character of these fields into account, one has:
$$
<\psi(u_1)\cdots \psi(u_n)\bar\psi(v_1)
\cdots\bar\psi(v_n)>\,=\,(-1)^{n(n-1)\over 2}\,\,
{\rm det}[\,{1\over u_i-v_j}\,]\,.
\eqn\apbquince
$$
By means of the so-called Cauchy determinant formula,
$$
{\rm det}[\,{1\over u_i-v_j}\,]\,=\,(-1)^{n(n-1)\over 2}\,\,
{\prod_{i<j}\,(u_i-u_j)\,(v_i-v_j)\over
\prod_{i,j}\,(u_i-v_j)}\,,
\eqn\apbdseis
$$
the correlator appearing in \apbonce\ can be written as:
$$
<\psi(t_1)\cdots \psi(t_n)\bar\psi(t_{n+1})
\cdots\bar\psi(t_{2n})>\,=\,
{\prod_{i<j}^n\,\,(t_i-t_j)^2\,(t_{n+i}-t_{n+j})^2\over
\prod_{i<j}^{2n}\,\,(t_i-t_j)}\,.
\eqn\apbdsiete
$$
Inserting the value given in eq. \apbdsiete\ for the
fermionic vacuum expectation value into the definition of 
$J_{2n}^{m}\,(a\,,b\,,\rho\,)$, it is easy to realize
that for the particular value $\rho\,=\,{1\over 2}$ the
$2n$-dimensional integrals \apbonce\ reduce to the
product of two $n$-dimensional integrals of the type
\apbdos. Actually, one has:
$$
J_{2n}^{m}\,(a\,,b\,,{1\over 2}\,)\,=\,
I_{n}^{0}\,(a\,,b\,,1\,)\,I_{n}^{m}\,(a\,,b\,,1\,)\,.
\eqn\apbdocho
$$
It can be easily proved that our result \apbdoce\
satisfies \apbdocho. In fact what happens is that for
$\rho\,=\,{1\over 2}$ the factors in \apbdoce\ with even
(odd) product index $i$ give rise to the function 
$I_{n}^{0}\,(a\,,b\,,1\,)$ ($I_{n}^{m}\,(a\,,b\,,1\,)$
respectively). On the other hand the integrals  
$J_{2n}^m\,(a,b,\rho)$ satisfy a recursion relation
similar to the one satisfied by the functions 
$I_{n}^m\,(a,b,\rho)$. Proceeding as in the derivation of
eq. \apbcinco, we get:
$$
J_{2n}^m\,(a,b,\rho)\,=\,\sum_{p=0}^m\,(-1)^{m-p}\,
{p+n-m\choose p}\,J_{2n}^{m-p}\,(b,a,\rho)\,.
\eqn\apbdnueve
$$
It is not difficult to prove that our ansatz \apbdoce\
satisfies the relation \apbdnueve. As  happened with eq.
\apbtres, this fact is a highly non-trivial check of eq.
\apbdoce.

Closely related to the functions
$J_{2n}^{m}\,(a\,,b\,,\rho\,)$ are the integrals:
$$
\eqalign{
\tilde J_{2n}^{m}\,(a\,,b\,,\rho\,)\,\equiv&\,\oint\,
\{\,[\,(t_{1}\cdots t_{m})^{a+1}\,
(t_{m+1}\cdots t_{n})^{a}\,
+\,{\rm  permutations}\,]\,\,(t_{n+1}\cdots t_{2n})^{a+1}\,\times \cr
&\times\,<\psi(t_1)\cdots \psi(t_n)\bar\psi(t_{n+1})
\cdots\bar\psi(t_{2n})>\,+\,{\rm  permutations}\,\}\times\cr
&\times\,\prod_{i=1}^{2n} (1-t_i)^b
\prod_{i<j}^{2n}\,(\,t_i\,-\,t_j\,)^{2\rho}\,,\cr}
\eqn\apbveinte
$$
where again $0\le m\le n$. In 
$\tilde J_{2n}^{m}\,(a\,,b\,,\rho\,)$ $n-m$ variables
chosen among the $n$ arguments of the fields $\psi$ have
an exponent which is one unit lower than the exponents of
the remaining variables. One can also write down a closed
expression for these multiple integrals:
$$
\eqalign{
\tilde J_{2n}^{m}\,(a\,,b\,,\rho\,)\,=&\,{n\choose m}\mu_{2n}(\rho)\,
\prod_{i=0}^{2n-2m-1}\,
{\Gamma(1+a+i(\rho-{1\over 2})+[{i+1\over 2}])\,
\Gamma(1+b+i(\rho-{1\over 2})+[{i\over 2}])\over
\Gamma(1+a+b+n+(\rho-{1\over 2})(2n-1+i)+[{i+1\over 2}])}\times\cr\cr
&\times\prod_{i=2n-2m}^{2n-1}\,
{\Gamma(2+a+i(\rho-{1\over 2})+[{i\over 2}])\,
\Gamma(1+b+i(\rho-{1\over 2})+[{i\over 2}])\over
\Gamma(2+a+b+n+(\rho-{1\over 2})(2n-1+i)+[{i\over 2}])}\,.
\cr}
\eqn\apbvuno
$$
Let us present the arguments in support of the result
\apbvuno. First of all, for the extreme values of $m$
(\ie\ for $m=0$ and $m=n$),   
$\tilde J_{2n}^{m}$ reduce to the previously studied
functions $ J_{2n}^{m}$. In fact, by inspecting the
definitions of these two types of integrals (eqs.
\apbonce\ and \apbveinte),  one easily concludes that:

$$
\eqalign{
\tilde J_{2n}^{\,0}\,(a\,,b\,,\rho\,)\,=&
J_{2n}^{\,n}\,(a\,,b\,,\rho\,)\cr\cr
\tilde J_{2n}^{\,n}\,(a\,,b\,,\rho\,)\,=&
J_{2n}^{\,0}\,(a+1\,,b\,,\rho\,)\,.
\cr}
\eqn\apbvdos
$$
Secondly, for $\rho\,=\,{1\over 2}$, the integrals
\apbveinte\ can be put in terms of the functions $I_n^m$
at $\rho\,=\,1$ :
$$
\tilde J_{2n}^{m}\,(a\,,b\,,{1\over 2}\,)\,=\,
I_{n}^{n}\,(a\,,b\,,1\,)\,I_{n}^{m}\,(a\,,b\,,1\,)\,.
\eqn\apbvtres
$$
Moreover, for $m=n-1$ the value of the right-hand side of 
\apbveinte\ can be given in terms of the known function 
$\tilde J_{2n}^{n}$:
$$
\tilde J_{2n}^{n-1}\,(a\,,b\,,\rho\,)\,=\,n\,\,\,
{{3\over 2}+a+b+(2n-1)\rho\over 1+a}\,\,\,
\tilde J_{2n}^{\,n}\,(b-1\,,a+1\,,\rho\,)\,.
\eqn\apbvcuatro
$$
The result \apbvcuatro\ can be obtained by employing the
same method used to derive \apbdiez. In can be easily
verified that the expression \apbvuno\ satisfies eqs.
\apbvdos-\apbvcuatro. It is important to point out that,
although the integrals $\tilde J_{2n}^{m}$ do not appear
directly in our calculation of the ${\rm osp}(1|2)$
structure constants, they are needed in some intermediate
steps (see below).

We shall also need integrals where the powers of some
of the factors $1-t_i$ are lowered in one unit. For
illustrative purposes, let us first consider  the case in
which there is no fermionic correlator in the integrand. 
We define the integrals 
${\cal I}_{2n}^{\,m}\,(a,b,\rho)$ with $0\le m\le n$ by
means of the expression:
$$
\eqalign{
{\cal I}_{2n}^{\,m}\,(a,b,\rho)\,=&\,\oint\,\{\,
\prod_{i=1}^{n}\,t_i^{a}\,t_{n+i}^{a+1}\,\,[\,
\prod_{i=1}^{m}(1-t_{n+i})^b\,\prod_{i=m+1}^{n}
(1-t_{n+i})^{b-1}\,+\,
{\rm permutations}\,]\times\cr
&\times\prod_{i=1}^{n}\,(1-t_{i})^b\,+\,
{\rm permutations}\,\}
\prod_{i<j}^{2n}\,(t_i-t_j)^{2\rho}\,.
\cr}
\eqn\apbvcinco
$$
For low values of $n$,  the  ${\cal I}_{2n}^{\,m}\,$'s 
 can be given in terms of our
previous results. From these particular cases one can easily
guess the general form of these functions. One expects now to
have products of $\Gamma$-functions similar to the ones in
\apbtres,  where now, in addition, 
$b$ is shifted in some of the arguments of the
$\Gamma$'s. In fact, the general expression of 
${\cal I}_{2n}^{\,m}\,(a,b,\rho)$ is given by:
$$
\eqalign{
{\cal I}_{2n}^{\,m}\,&(a,b,\rho)\,=\,
{2n \choose n}\,
{n \choose m}\,\lambda_{2n}(\rho)\,
\prod_{i=0}^{n-m-1}\,
{\Gamma (1+a+i\rho)\,\Gamma (b+i\rho)\over
\Gamma (2+a+b+(2n-1+i)\rho)}\,\times\cr
&\times\prod_{i=n-m}^{n-1}\,
{\Gamma (1+a+i\rho)\,\Gamma (1+b+i\rho)\over
\Gamma (2+a+b+(2n-1+i)\rho)}\,
\prod_{i=n}^{2n-m-1}\,
{\Gamma (2+a+i\rho)\,\Gamma (1+b+i\rho)\over
\Gamma (2+a+b+(2n-1+i)\rho)}\,\times\cr
&\times\prod_{i=2n-m}^{2n-1}\,
{\Gamma (2+a+i\rho)\,\Gamma (1+b+i\rho)\over
\Gamma (3+a+b+(2n-1+i)\rho)}\,.
\cr}
\eqn\apbvseis
$$
It is an easy exercise to check that eq. \apbvseis\
gives the correct result in the decoupling point
$\rho=0$. Moreover, one can check  eq. \apbvseis\ for
the extreme values of $m$. Indeed, for $m=n$ one must have
(compare the definitions \apbdos\ and \apbvcinco):
$$
{\cal I}_{2n}^{\,n}\,(a,b,\rho)\,\,=\,\,
I_{2n}^n\,(a,b,\rho)\,.
\eqn\apbvsiete
$$
It is straightforward to verify that our solution
\apbvseis\ satisfies eq. \apbvsiete\ when 
$I_{2n}^n\,(a,b,\rho)$ is given by \apbtres. When $m=0$,
it is also possible to derive the form of the integrals
\apbvcinco\ for arbitrary $n$. Suppose that in the
definition of ${\cal I}_{2n}^{\,0}\,(a,b,\rho)$, 
$$
\eqalign{
{\cal I}_{2n}^{\,0}\,(a,b,\rho)\,\,=&\,\,\oint\,\{\,
\prod_{i=1}^n\,t_{i}^{a}\,t_{n+i}^{a+1}\,
(1-t_{i})^{b}\,(1-t_{n+i})^{b-1}\,+\,
{\rm permutations}\,\}\times\cr
&\times\prod_{i<j}^{2n}\,(t_i-t_j)^{2\rho}\,,\cr}
\eqn\apbvocho
$$
we substitute the identity 
$(1-t_{i})^{b}\,=\,(1-t_{i})^{b-1}\,(1-t_{i})$ in all the
$1-t_i$ factors raised to the power $b$. The resulting
integrals are of the form \apbdos\ and, actually, one has:
$$
{\cal I}_{2n}^{\,0}\,(a,b,\rho)\,\,=\,\sum_{l=0}^n\,
(-1)^l\,{n+l\choose n}\,\,
I_{2n}^{n+l}(a,b-1,\rho)\,, 
\eqn\apbvnueve
$$
which gives ${\cal I}_{2n}^{\,0}\,(a,b,\rho)$ in terms of
known quantities. Let us prove that the value of 
${\cal I}_{2n}^{\,0}\,(a,b,\rho)$ obtained from the
right-hand side of \apbvnueve\ is equal to the one
displayed in eq. \apbvseis. The explicit value of 
$I_{2n}^{n+l}(a,b-1,\rho)$ is (see eq. \apbtres):
$$
\eqalign{
I_{2n}^{n+l}\,(a,b-1,\rho)\,=&\,{2n \choose n+l}\,\lambda_{2n}(\rho)\,
\prod_{i=0}^{n-l-1}\,
{\Gamma (1+a+i\rho)\,\Gamma (b+i\rho)\over
\Gamma (1+a+b+(2n-1+i)\rho)}\,\times\cr
&\times \prod_{i=n-l}^{2n-1}\,
{\Gamma (2+a+i\rho)\,\Gamma (b+i\rho)\over
\Gamma (2+a+b+(2n-1+i)\rho)}\,.\cr}
\eqn\apbtreinta
$$
Using the property $\Gamma(1+x)=x\Gamma(x)$ in eq.
\apbtreinta, one can extract the $l$-dependent part of 
$I_{2n}^{n+l}(a,b-1,\rho)$ as follows:
$$
I_{2n}^{n+l}\,(a,b-1,\rho)\,=\,{2n \choose n+l}\,
C_n^l(a,b,\rho)\,\Omega_{2n}\,(a,b,\rho)\,.
\eqn\apbtuno
$$
In eq. \apbtuno\ $C_n^l(a,b,\rho)$ is given by:
$$
C_n^l(a,b,\rho)\,=\,
\prod_{i=0}^{n-l-1}\,[1+a+b+(2n-1+i)\rho]\,\,
\prod_{i=n-l}^{n-1}\,(1+a+i\rho)\,,
\eqn\apbtdos
$$
while $\Omega_{2n}\,(a,b,\rho)$ denotes the quantity:
$$
\eqalign{
\Omega_{2n}\,(a,b,\rho)\,=&\,\lambda_{2n}(\rho)\,
\prod_{i=0}^{n-1}\,
{\Gamma (1+a+i\rho)\,\Gamma (b+i\rho)\over
\Gamma (2+a+b+(2n-1+i)\rho)}\,\times\cr
&\times \prod_{i=n}^{2n-1}\,
{\Gamma (2+a+i\rho)\,\Gamma (b+i\rho)\over
\Gamma (2+a+b+(2n-1+i)\rho)}\,.\cr}
\eqn\apbttres
$$
Amazingly, the sum in \apbvnueve\ can be done explicitly
by means of the binomial identity \ocuatro. One has:
$$
\sum_{l=0}^n\,(-1)^l\,{n+l\choose n}\,{2n\choose n+l}
\,C_n^l(a,b,\rho)\,=\,
{2n\choose n}\,\prod_{i=n}^{2n-1}\,(b+i\rho)\,.
\eqn\apbtcuatro
$$
Therefore the expression of 
${\cal I}_{2n}^{\,0}\,(a,b,\rho)$ that we get is:

$$
{\cal I}_{2n}^{\,0}\,(a,b,\rho)\,\,=\,\,
{2n\choose n}\,\Bigl(\,\,
\prod_{i=n}^{2n-1}\,(b+i\rho)\,\,\Bigr)\,
\Omega_{2n}\,(a,b,\rho)\,.
\eqn\apbtcinco
$$
Using again $\Gamma(1+x)=x\Gamma(x)$, one easily proves
that the right-hand side of eq. \apbtcinco\ equals the
value given by eq. \apbvseis\ for $m=0$. 

The same methodology that we have applied to obtain eq.
\apbvseis\ can be used to get the values of the integrals:
$$
\eqalign{
{\cal J}_{2n}^{\, m}(a,b,\rho)\,\equiv&\,\,\,\oint\,\{\,
\prod_{i=1}^n\,t_{i}^{a}\,t_{n+i}^{a+1}\,
(1-t_{i})^{b}\,[\,\prod_{i=1}^{m}\,(1-t_{n+i})^{b}\,
\prod_{i=m+1}^{n}\,(1-t_{n+i})^{b-1}\,+\,
{\rm permutations}\,]\times\cr
&\times\,<\psi(t_1)\cdots \psi(t_n)\bar\psi(t_{n+1})
\cdots\bar\psi(t_{2n})>\,+\,{\rm  permutations}\,\}\times\cr
&\times\prod_{i<j}^{2n}\,(\,t_i\,-\,t_j\,)^{2\rho}\,,
\cr}
\eqn\apbtseis
$$
where $0\le m\le n$. Notice that the only difference
between the definitions \apbvcinco\ and \apbtseis\ is the
presence in the integrand of the latter of the fermionic
correlator. As we are now going to argue, the 
${\cal J}_{2n}^{\,m}\,$'s are given by:
$$
\eqalign{
{\cal J}_{2n}^{\,m}\,(a\,,b\,,\rho\,)\,
=&\,{n\choose m}\mu_{2n}(\rho)\,
\prod_{i=0}^{2n-2m-1}\,
{\Gamma(1+a+i(\rho-{1\over 2})+[{i+1\over 2}])\,
\Gamma(b+i(\rho-{1\over 2})+[{i+1\over 2}])\over
\Gamma(1+a+b+n+(\rho-{1\over 2})(2n-1+i)+
[{i\over 2}])}\times\cr\cr
&\times\prod_{i=2n-2m}^{2n-1}\,
{\Gamma(1+a+i(\rho-{1\over 2})+[{i+1\over 2}])\,
\Gamma(1+b+i(\rho-{1\over 2})+[{i\over 2}])\over
\Gamma(1+a+b+n+(\rho-{1\over 2})(2n-1+i)
+[{i+1\over 2}])}\,.
\cr}
\eqn\apbtsiete
$$
Indeed, eq. \apbtsiete, when $\rho={1\over 2}$, can be
written as:
$$
{\cal J}_{2n}^{\,m}\,(a\,,b\,,\,{1\over 2}\,)\,=\,
 I_{n}^{\,0}\,(a\,,b\,,\,1\,)\,\,
I_{n}^{\,m}\,(b-1\,,a+1\,,\,1\,)\,,
\eqn\apbtocho
$$
which is the result expected from the definition
\apbtseis. Moreover, for $m=n$ one should have:
$$
{\cal J}_{2n}^{\,n}\,(a\,,b\,,\rho\,)\,=\,
\tilde J_{2n}^{\,0}\,(a\,,b\,,\rho\,)\,=\,
J_{2n}^{\,n}\,(a\,,b\,,\rho\,)\,,
\eqn\apbtnueve
$$
and a simple inspection of eqs. \apbtsiete\ and \apbvuno\
shows that this is indeed the case. Finally, for $m=0$ the
analogue of eq. \apbvnueve\ is:
$$
{\cal J}_{2n}^{\,0}\,(a\,,b\,,\rho\,)\,=\,
\,\sum_{l=0}^n\,(-1)^l\,
\tilde
J_{2n}^{\,l}\,(a\,,b-1\,,\rho\,)\,.
\eqn\apbcuarenta 
$$
The sum of eq. \apbcuarenta\ can also be performed and
the result, similar to \apbtcinco, matches perfectly with
our ansatz \apbtsiete.

Let us now try to generalize the definitions \apbonce\ 
and \apbtseis\ to include the case in which the integrals
are performed in an odd-dimensional region. The obvious
difficulty to overcome in this case is the fact that the
correlator of an odd number of fermionic fields vanishes.
To solve this problem, we include in the correlator an
extra field $\bar\psi$  with its argument placed at
infinity. In fact, the definition of 
$J_{2n+1}^{m}\,(a\,,b\,,\rho\,)$ that we shall adopt is:
$$
\eqalign{
J&_{2n+1}^{m}\,(a\,,b\,,\rho\,)\,\equiv\,\cr
&\equiv\,-\oint\,\{\,(t_1\cdots t_{n+1})^a\,
[\,(t_{n+2}\cdots t_{n+m+1})^{a+1}\,
(t_{n+m+2}\cdots t_{2n+1})^{a}\,
+\,{\rm  permutations}\,]\times\cr
&\times\,{\rm lim}_{R\rightarrow\infty}
R<\psi(t_1)\cdots \psi(t_{n+1})\bar\psi(t_{n+2})
\cdots\bar\psi(t_{2n+1})\bar\psi(R)>\,
+\,{\rm permutations}\,\}\times\cr
&\times\,\prod_{i=1}^{2n+1} (1-t_i)^b
\prod_{i<j}^{2n+1}\,(\,t_i\,-\,t_j\,)^{2\rho}\,,\cr}
\eqn\apbcuno
$$
where we have multiplied  the correlator by $R$ in order
to get a non-vanishing result in the limit 
$R\rightarrow\infty$.  The minus sign in \apbcuno\ appears
because, for all the permutations, $\bar\psi(R)$ is
inserted to the right of all the other fields in the
correlator (\ie\ the field $\bar\psi(R)$ behaves as
an spectator). The final justification to define
$J_{2n+1}^{m}\,(a\,,b\,,\rho\,)$ as in eq. \apbcuno\ is
the fact that,  in our calculation of the ${\rm
osp}(1|2)$ structure constants,  these
integrals appear precisely in this form . Moreover, there is a
nice generalization of the eq. \apbdoce\ that gives the value
of $J_{N}^{m}\,(a\,,b\,,\rho\,)$ for any integer $N$ and 
$m\le [{N\over 2}]$. This expression is:
$$
\eqalign{
&J_{N}^{m}\,(a\,,b\,,\rho\,)\,=\cr
&=\,{[N/2]\choose m}\mu_{N}(\rho)\,
\prod_{i=0}^{2[N/2]-2m-1}\,
{\Gamma(1+a+i(\rho-{1\over 2})+[{i\over 2}])\,
\Gamma(1+b+i(\rho-{1\over 2})+[{i\over 2}])\over
\Gamma(1+a+b+(\rho-{1\over 2})(N-1+i)+[{i+N\over 2}])}\times\cr\cr
&\times\prod_{i=2[N/2]-2m}^{N-1}\,
{\Gamma(1+a+i(\rho-{1\over 2})+[{i+1\over 2}])\,
\Gamma(1+b+i(\rho-{1\over 2})+[{i\over 2}])\over
\Gamma(1+a+b+(\rho-{1\over 2})(N-1+i)
+[{i+N+1\over 2}])}\,.
\cr}
\eqn\apbcdos
$$
Notice that the different behaviour for even and odd $N$
is reproduced by putting  $N$ inside the
integer part symbol in eq. \apbcdos. Similarly, we can define 
${\cal J}_{N}^{\,m}\,(a\,,b\,,\rho\,)$ for odd $N$ as
follows:
$$
\eqalign{
{\cal J}_{2n+1}^{\, m}&(a,b,\rho)\,=-\,\,\,\oint\,\{\,
\prod_{i=1}^{n+1}\,t_{i}^{a}\,(1-t_{i})^{b}
\prod_{i=1}^{n}\,t_{n+i+1}^{a+1}\,\times\cr
&\times[\,\prod_{i=1}^{m}\,(1-t_{n+i+1})^{b}\,
\prod_{i=m+1}^{n}\,(1-t_{n+i+1})^{b-1}\,+\,
{\rm permutations}\,]\times\cr
&\times\,{\rm lim}_{R\rightarrow\infty}\,
R<\psi(t_1)\cdots \psi(t_{n+1})\bar\psi(t_{n+2})
\cdots\bar\psi(t_{2n+1})\bar\psi(R)>\,
+\,{\rm  permutations}\,\}\times\cr
&\times\prod_{i<j}^{2n+1}\,(\,t_i\,-\,t_j\,)^{2\rho}\,,
\cr}
\eqn\apbctres
$$
and the same rule used to pass from eq. \apbdoce\ to
\apbcdos\ serves to generalize \apbtsiete. If 
$m\le [{N\over 2}]$ one has:
$$
\eqalign{
{\cal J}_{N}^{\,m}&\,(a\,,b\,,\rho\,)\,=\cr
&=\,{[N/2]\choose m}\mu_{N}(\rho)\, 
\prod_{i=0}^{2[N/2]-2m-1}\,
{\Gamma(1+a+i(\rho-{1\over 2})+[{i+1\over 2}])\,
\Gamma(b+i(\rho-{1\over 2})+[{i+1\over 2}])\over
\Gamma(1+a+b+(\rho-{1\over 2})(N-1+i)+[{i+N\over 2}])}\times\cr\cr
&\times\prod_{i=2[N/2]-2m}^{N-1}\,
{\Gamma(1+a+i(\rho-{1\over 2})+[{i+1\over 2}])\,
\Gamma(1+b+i(\rho-{1\over 2})+[{i\over 2}])\over
\Gamma(1+a+b+(\rho-{1\over 2})(N-1+i)
+[{i+N+1\over 2}])}\,.
\cr}
\eqn\apbccuatro
$$

Let us finally point out that we have numerically checked,
for low dimensions, the values of the integrals given in
this Appendix.

\refout

\end